\documentclass[12pt,a4paper]{article}
\pdfoutput = 1
\usepackage{amsmath}
\usepackage{cite}
\usepackage{hyperref}
\usepackage{graphicx}
\usepackage{booktabs}

\newcommand{\Kstar}{\kern 0.2em\overline{\kern -0.2em K}^{*0}}

\begin{document}

\begin{titlepage}

\vspace*{-0.0truecm}

\begin{flushright}
Nikhef-2014-031
\end{flushright}

\vspace*{1.3truecm}

\begin{center}
\boldmath
{\Large{\bf 
A Roadmap to Control Penguin Effects in $B^0_d\to J/\psi K_{\rm S}^0$ and $B^0_s\to J/\psi \phi$}}
\unboldmath
\end{center}

\vspace{0.9truecm}

\begin{center}
{\bf Kristof De Bruyn\,${}^a$ and Robert Fleischer\,${}^{a,b}$}

\vspace{0.5truecm}

${}^a${\sl Nikhef, Science Park 105, NL-1098 XG Amsterdam, Netherlands}

${}^b${\sl  Department of Physics and Astronomy, Vrije Universiteit Amsterdam,\\
NL-1081 HV Amsterdam, Netherlands}

\end{center}

\vspace{1.4cm}
\begin{abstract}
\vspace{0.2cm}\noindent
Measurements of CP violation in $B^0_d\to J/\psi K_{\rm S}^0$ and $B^0_s\to J/\psi \phi$
decays play key roles in testing the quark-flavour sector of the Standard Model. The
theoretical interpretation of the corresponding observables is limited by uncertainties from
doubly Cabibbo-suppressed penguin topologies. With continuously increasing experimental
precision, it is mandatory to get a handle on these contributions, which cannot be calculated
reliably in QCD. In the case of the measurement of $\sin2\beta$ from $B^0_d\to J/\psi K_{\rm S}^0$,
the $U$-spin-related decay $B^0_s\to J/\psi K_{\rm S}^0$ offers a tool to control the penguin 
effects. As the required measurements are not yet available, we use data for decays with
similar dynamics and the $SU(3)$ flavour symmetry to constrain the size of the expected 
penguin corrections. We predict the CP asymmetries of $B^0_s\to J/\psi K_{\rm S}^0$ and
present a scenario to fully exploit the physics potential of this decay, emphasising also the 
determination of hadronic parameters and their comparison with theory. In the case 
of the benchmark mode $B^0_s\to J/\psi \phi$ used to determine the $B^0_s$--$\bar B^0_s$ mixing 
phase $\phi_s$ the penguin effects can be controlled through $B^0_d\to J/\psi \rho^0$ and 
$B^0_s\to J/\psi \Kstar$ decays. The LHCb collaboration has recently presented pioneering 
results on this topic. We analyse their implications and present a roadmap for 
controlling the penguin effects.
\end{abstract}

\vspace*{0.5truecm}
\vfill
\noindent
December 2014
\vspace*{0.5truecm}

\end{titlepage}

\thispagestyle{empty}
\vbox{}
\newpage

\setcounter{page}{1}

\section{Introduction}\label{sec:intro}
The data of the first run of the Large Hadron Collider (LHC) at CERN have led to the exciting 
discovery of the Higgs boson \cite{Higgs-ATLAS,Higgs-CMS}
and are, within the current level of precision, globally consistent
with the picture of the Standard Model (SM). The next run of the LHC at almost the double 
centre-of-mass energy of the colliding protons, which will start in spring 2015, will open various 
new opportunities in the search for New Physics (NP) \cite{Ellis}. These will be both in the form of 
direct searches for new particles at the ATLAS and CMS experiments, and in the form of
high-precision analyses of flavour physics observables at the LHCb experiment.
Concerning the latter avenue, also the Belle II experiment at the KEK $e^+e^-$ Super $B$ 
Factory will enter the stage in the near future \cite{Abe:2010gxa}. 
The current LHC data suggest that we have to 
prepare ourselves to deal with smallish NP effects, and it thus becomes mandatory to 
have a critical look at the theoretical assumptions underlying the experimental analyses.

Concerning measurements of CP violation, the $B^0_d\to J/\psi K_{\rm S}^0$ and 
$B^0_s\to J/\psi \phi$ decays play outstanding roles as they allow determinations 
of the $B_q^0$--$\bar{B}_q^0$ mixing phases $\phi_d$ and $\phi_s$, respectively.
These quantities take the forms
\begin{equation}\label{mix-phases}
\phi_d=2\beta + \phi_d^{\rm NP}, \quad \phi_s= -2\lambda^2\eta + \phi_s^{\rm NP},
\end{equation} 
where $\beta$ is the usual angle of the unitarity triangle (UT) of the Cabibbo--Kobayashi--Maskawa
(CKM) matrix \cite{cab,KM} and 
\begin{equation}\label{phis-SM}
\phi_s^{\rm SM}=-2\lambda^2\eta=-(2.086^{+0.080}_{-0.069})^{\circ}
\end{equation}
in the SM  \cite{Charles:2011va}. The $\lambda$ and $\eta$ are two of the
Wolfenstein parameters \cite{wolf} of the CKM matrix. The CP-violating phases 
$\phi_q^{\rm NP}$, which vanish in the SM, allow for NP contributions entering 
through $B^0_q$--$\bar B^0_q$ mixing.

The theoretical precision for the extraction of $\phi_d$ and $\phi_s$ from the CP asymmetries 
of the $B^0_d\to J/\psi K_{\rm S}^0$ and $B^0_s\to J/\psi \phi$ decays  is 
limited by doubly Cabibbo-suppressed penguin contributions. The corresponding non-perturbative 
hadronic parameters cannot be calculated in a reliable way within QCD. However, in the era 
of high-precision measurements, these effects have to be controlled with the final goal to match 
the experimental and theoretical precisions  
\cite{RF-psiK,RF-ang,RF-B99,CPS,FFJM,FFM,MJ,LWX}. 

As was pointed out in Ref.~\cite{RF-psiK}, $B^0_s\to J/\psi K_{\rm S}^0$ is 
related to $B^0_d\to J/\psi K_{\rm S}^0$ through the $U$-spin symmetry of strong interactions, 
and allows a determination of the penguin corrections to the measurement of $\phi_d$. 
Concerning the $B^0_s\to J/\psi \phi$ channel, an analysis of
CP violation is more involved as the final state 
consists of two vector mesons and thus is a mixture of different CP eigenstates which have to 
be disentangled through an angular analysis of their decay products \cite{DDLR,DDF}. In this 
case, the decays $B^0_d\to J/\psi \rho^0$ \cite{RF-ang} and $B^0_s\to J/\psi \Kstar$ 
\cite{FFM} are tools to take the penguin effects into account. The LHCb collaboration
has very recently presented the first polarisation-dependent measurements of $\phi_s$
from $B^0_s\to J/\psi \phi$ in Ref.~\cite{Aaij:2014zsa}. We shall discuss the implications
of these exciting new results in detail. 

Since a measurement of CP violation in $B^0_s\to J/\psi K_{\rm S}^0$ is not yet available, 
we use the $SU(3)$ flavour symmetry and plausible assumptions for various
modes of similar decay dynamics to constrain the relevant penguin parameters. Following these
lines, we assess their impact on the measurement of $\phi_d$ and predict the 
CP-violating observables of $B^0_s\to J/\psi K_{\rm S}^0$. In our benchmark scenario, we
discuss also the determination of CP-conserving strong amplitudes, which will provide valuable
insights into non-factorisable $U$-spin-breaking effects through the comparison with 
theoretical form-factor calculations.

Concerning the $B^0_s\to J/\psi \Kstar$ channel, measurements of CP 
violation are also not yet available. However, in the case of $B^0_d\to J/\psi \rho^0$, 
the LHCb collaboration has recently announced the first results of a pioneering 
study \cite{LHCb-psi-rho}, presenting in particular a measurement of mixing-induced CP 
violation and constraints on the penguin effects. This new experimental development 
was made possible through the implementation of the method proposed by Zhang and 
Stone in Ref.~\cite{ZS}. 
We shall have a detailed look at these exciting measurements and discuss important differences 
between the penguin probes $B^0_d\to J/\psi \rho^0$ and $B^0_s\to J/\psi \Kstar$. We extract
hadronic parameters from the $B^0_d\to J/\psi \rho^0$ data, allowing insights into 
$SU(3)$-breaking and non-factorisable effects through a comparison with theory, and 
point out a new way to combine the information provided by the $B^0_s\to J/\psi \Kstar$,
$B^0_d\to J/\psi \rho^0$ system in a global analysis of the $B^0_s\to J/\psi \phi$ 
penguin parameters. 

The outline of this paper is as follows: in Section~\ref{sec:hadr}, we introduce the general 
formalism to deal with the penguin effects. In Section~\ref{sec:psiP}, we explore 
the constraints of the currently available data for the penguin contributions to the 
$B^0_{d,s}\to J/\psi K_{\rm S}^0$ system, while we turn to the discussion of the most
recent LHCb results for $B^0_s\to J/\psi \phi$ and the penguin probes $B^0_d\to J/\psi \rho^0$,
$B^0_s\to J/\psi \Kstar$ in Section~\ref{sec:psiV}. In Section~\ref{sec:road}, 
we outline a roadmap for dealing with the hadronic penguin uncertainties in the determination 
of $\phi_d$ and $\phi_s$. Finally, we summarise our conclusions in Section~\ref{sec:concl}.

\section{CP Violation and Hadronic Penguin Shifts}\label{sec:hadr}
For the neutral $B_q$ decays ($q=d,s$) discussed in this paper, the transition amplitudes 
can be written in the following form \cite{RF-ang}:
\begin{align}
A(B^0_q\to f)\equiv A_f =\: & \phantom{\eta_f}{\cal N}_f\left[1-
b_f e^{\rho_f}e^{+i\gamma}\right]\:, \label{ampl}\\
A(\bar B^0_q\to f)\equiv \bar A_f =\: & \eta_f{\cal N}_f\left[1-
b_f e^{\rho_f}e^{-i\gamma}\right]\:.\label{ampl-CP}
\end{align}
Here $\eta_f$ is the CP eigenvalue of the final state $f$, ${\cal N}_f$ is a CP-conserving
normalisation factor representing the dominant tree topology, $b_f$ parametrises the relative 
contribution from the penguin topologies, $\rho_f$ is the CP-conserving strong phase difference 
between the tree and penguin contributions, whereas their relative weak phase is given by the UT 
angle $\gamma$. The parameters ${\cal N}_f$ and $b_f$ depend both on CKM factors and on 
hadronic matrix elements of four-quark operators entering the corresponding low-energy 
effective Hamiltonian.

In order to extract information on $\phi_q$, CP-violating asymmetries are measured \cite{RF-rev}:
\begin{equation}\label{ACP}
\frac{|A(B^0_q(t)\to f)|^2-|A(\bar B^0_q(t)\to f)|^2}{|A(B^0_q(t)\to f)|^2-|A(\bar B^0_q(t)\to f)|^2}=
\frac{{\cal }{\cal A}_{\rm CP}^{\rm dir}
\cos(\Delta M_qt)+{\cal A}_{\rm CP}^{\rm mix}\sin(\Delta M_qt)}{\cosh(\Delta\Gamma_qt/2)+
{\cal A}_{\Delta\Gamma}\sinh(\Delta\Gamma_qt/2)}\:,
\end{equation} 
where the dependence on the decay time $t$ enters through $B^0_q$--$\bar B^0_q$ oscillations, 
and \mbox{$\Delta M_q\equiv M^{(q)}_{\rm H}-M^{(q)}_{\rm L}$} and 
\mbox{$\Delta\Gamma_q\equiv \Gamma_{\rm L}^{(q)}-\Gamma_{\rm H}^{(q)}$} denote
the mass and decay width differences of the two $B_q$ mass eigenstates,  respectively.

Using Eqs.~\eqref{ampl} and \eqref{ampl-CP}, the direct and mixing-induced CP asymmetries 
${\cal A}_{\rm CP}^{\rm dir}$ and ${\cal A}_{\rm CP}^{\rm mix}$ take 
the following forms \cite{RF-ang}:\footnote{Whenever information from both 
$B_q^0\to f$ and $\bar{B}_q^0\to f$ decays is needed to determine an 
observable, as is the case for CP asymmetries or untagged branching ratios, 
we use the notation $B_d$ and $B_s$.}
 \begin{equation}\label{AD-expr}
 {\cal A}_{\rm CP}^{\rm dir}(B_q\to f)=\frac{2 b_f \sin\rho_f\sin\gamma}{1-
 2b_f\cos\rho_f\cos\gamma+b_f^2}\:,
 \end{equation}
\begin{equation}\label{AM-expr}
 {\cal A}_{\rm CP}^{\rm mix}(B_q\to f)=\eta_f\left[
 \frac{\sin\phi_q-2 b_f \cos\rho_f\sin(\phi_q+\gamma)+b_f^2\sin(\phi_q+2\gamma)}{1-
 2b_f\cos\rho_f\cos\gamma+b_f^2}\right]\:,
 \end{equation}
 while the observable $\mathcal{A}_{\Delta\Gamma}$ is given by
 \begin{equation}\label{Eq:ADG}
\mathcal{A}_{\Delta\Gamma}(B_q\to f)=-\eta_f\left[
 \frac{\cos\phi_q-2 b_f \cos\rho_f\cos(\phi_q+\gamma)+b_f^2\cos(\phi_q+2\gamma)}{1-
 2b_f\cos\rho_f\cos\gamma+b_f^2}\right]\:.
\end{equation}
 For the discussion of the penguin effects, the following expression will be particularly 
 useful (generalising the formulae given in Ref.~\cite{FFM}):
 \begin{equation}\label{phiq-eff-def}
 \frac{\eta_f {\cal A}_{\rm CP}^{\rm mix}(B_q\to f)}{\sqrt{1- 
 \left({\cal A}_{\rm CP}^{\rm dir}(B_q\to f)\right)^2}}=\sin(\phi_q+\Delta\phi_q^f) \equiv
 \sin(\phi_{q,f}^{\rm eff})\:,
 \end{equation}
where 
\begin{align}
\sin \Delta\phi_q^f & =\frac{-2b_f\cos\rho_f\sin\gamma+b_f^2\sin2\gamma}{\left(1-
2b_f\cos\rho_f\cos\gamma+b_f^2\right)\sqrt{1- \left({\cal A}_{\rm CP}^{\rm dir}(B\to f)\right)^2}}\:,\\
\cos \Delta\phi_q^f & =\frac{1-2b_f\cos\rho_f\cos\gamma+b_f^2\cos2\gamma}{\left(1-
2b_f\cos\rho_f\cos\gamma+b_f^2\right)\sqrt{1- \left({\cal A}_{\rm CP}^{\rm dir}(B\to f)\right)^2}}\:,
\end{align}
yielding
\begin{equation}\label{Eq:DeltaPhi_Def}
\tan  \Delta\phi_q^f = -\left[ \frac{2b_f\cos\rho_f\sin\gamma-b^2\sin2\gamma}{1-
2b_f\cos\rho_f\cos\gamma+b_f^2\cos2\gamma} \right]\:.
\end{equation}
It should be emphasised that $\Delta\phi_q^f$ is a phase shift which depends on the 
non-perturbative parameters $b_f$ and $\rho_f$ and cannot be calculated reliably
within QCD. In the case of $b_f=0$, the following simple situation arises:
\begin{equation}\label{AM-ideal}
{\cal A}_{\rm CP}^{\rm dir}(B_q\to f) |_{b_f=0} = 0\:, \qquad
\eta_f {\cal A}_{\rm CP}^{\rm mix}(B_q\to f) |_{b_f=0} = \sin \phi_q\:,
\end{equation}
allowing us to determine $\phi_q$ directly from the mixing-induced CP asymmetry.

Since in the decays $B^0_d\to J/\psi K_{\rm S}^0$ and $B^0_s\to J/\psi \phi$ the parameters 
corresponding to $b_f$ are doubly Cabibbo-suppressed, Eq.~\eqref{AM-ideal} 
is approximately valid. However, in the era of high-precision studies of 
CP violation, we nonetheless have to control these effects. As the corresponding penguin 
parameters are Cabibbo-allowed in the $B^0_s\to J/\psi K_{\rm S}^0$ and $B^0_d\to J/\psi\rho^0$, 
$B^0_s\to J/\psi \Kstar$ decays, these modes allow us to probe the penguin effects.
Making use of the $SU(3)$ flavour symmetry, we may subsequently convert the penguin 
parameters into their $B^0_d\to J/\psi K_{\rm S}^0$ and $B^0_s\to J/\psi \phi$ counterparts, where
in the latter case also plausible dynamical assumptions beyond the $SU(3)$ are
required.

\boldmath
\section{The $B^0_d\to J/\psi K_{\rm S}^0$, $B^0_s\to J/\psi K_{\rm S}^0$
System}\label{sec:psiP}
\unboldmath
\subsection{Decay Amplitudes and CP Violation}
%
%
%
\begin{figure}[t]
\center
\includegraphics[width=0.45\textwidth]{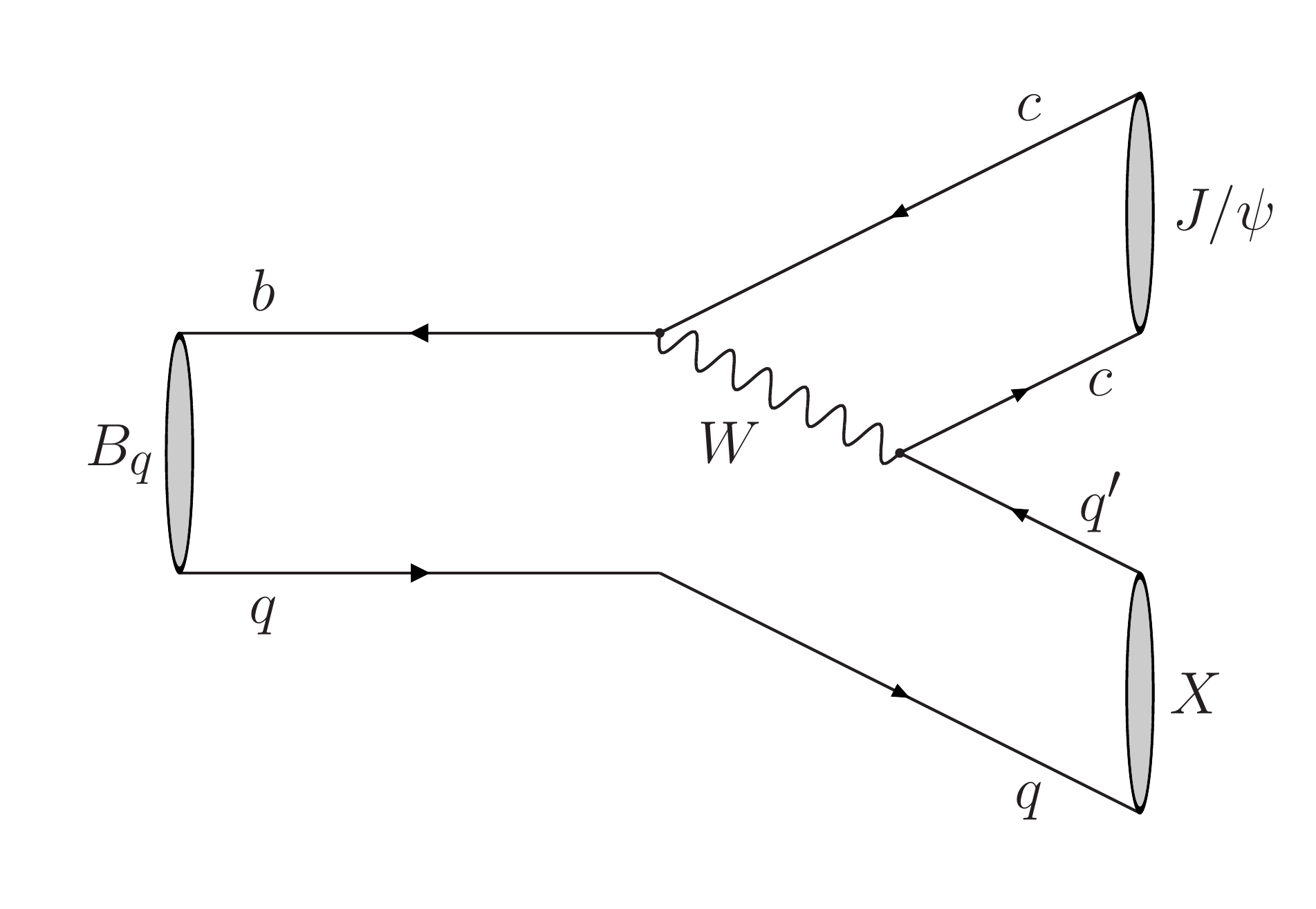}
\hfill
\includegraphics[width=0.45\textwidth]{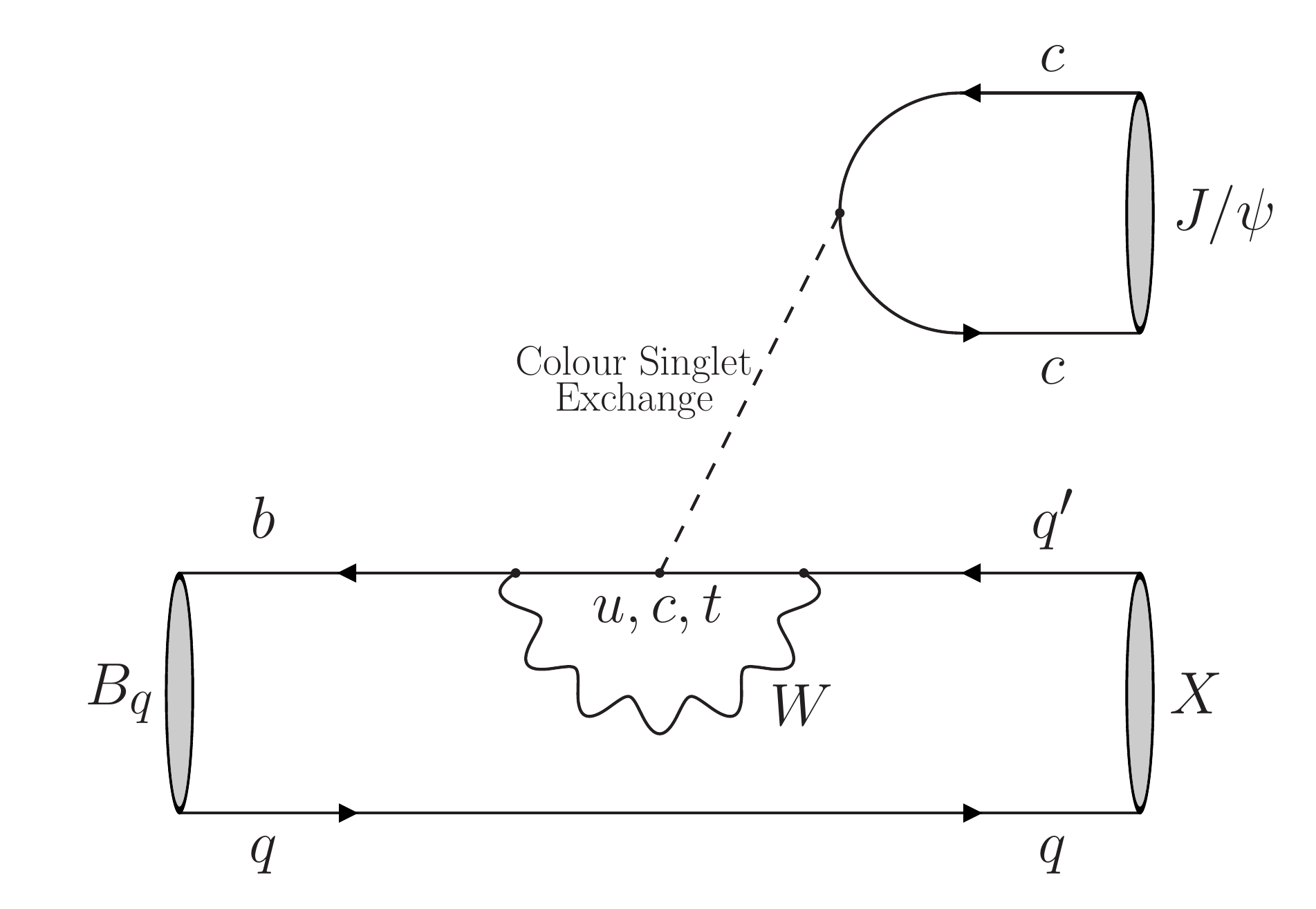}
\caption{Illustration of tree (left) and penguin (right) topologies contributing to the 
$B_q\to J/\psi X$ channels, where $q\in\{u,d,s\}$, $q'\in\{d,s\}$ and $X$ represents
any of the $\pi^0$, $\pi^+$, $K^+$, $K_{\rm S}^0$, $\rho^0$, $\phi^0$ or $\Kstar$ mesons.}
\label{Fig:Feynman_Tree}
\end{figure}

In the SM, the decay $B^0_d\to J/\psi K_{\rm S}^0$ into a CP eigenstate 
with eigenvalue $\eta_{J/\psi K_{\rm S}^0}=-1$ originates from a colour-suppressed
tree contribution and penguin topologies with $q$-quark exchanges ($q=u,c,t$), which are
described by CP-conserving amplitudes $C'$ and $P^{\prime(q)}$, respectively, and 
illustrated in Fig.~\ref{Fig:Feynman_Tree}.
The primes are introduced to remind us that we are dealing with 
a $\bar b\to \bar s c \bar c$ quark-level process. Using the unitarity 
of the CKM matrix, the $B^0_d\to J/\psi K_{\rm S}^0$ decay amplitude 
can be expressed in the following form \cite{RF-psiK}:
\begin{equation}\label{Eq:DA_Bd2JpsiKS}
A\left(B_d^0\rightarrow J/\psi K_{\mathrm S}^0\right) = \left(1-\frac{\lambda^2}{2}\right)\mathcal{A}'
\left[1+\epsilon a'e^{i\theta'}e^{i\gamma}\right]\:,
\end{equation}
where
\begin{equation}\label{Aprime-def}
\mathcal{A}' \equiv \lambda^2 A \left[C'+P^{\prime(c)}-P^{\prime(t)}\right]
\end{equation}
and
\begin{equation}\label{Eq:Penguin_Def}
a'e^{i\theta'} \equiv R_b\left[\frac{P^{\prime(u)}-P^{\prime(t)}}{C'+P^{\prime(c)}-P^{\prime(t)}}\right]
\end{equation}
are CP-conserving hadronic parameters. The Wolfenstein parameter $\lambda$ takes the 
value $\lambda \equiv |V_{us}|= 0.22551 \pm 0.00068$ \cite{Charles:2011va}, and
\begin{equation}
\epsilon \equiv \frac{\lambda^2}{1-\lambda^2}\:,\qquad
A \equiv \frac{|V_{cb}|}{\lambda^2}\:,\qquad
R_b \equiv \left(1-\frac{\lambda^2}{2}\right)\frac{1}{\lambda}\left|\frac{V_{ub}}{V_{cb}}\right|
\end{equation}
are combinations of CKM matrix elements. The parameter $a'$ measures the size of the penguin
topologies with respect to the tree contribution, and is associated with the CP-conserving
strong phase $\theta'$.  A key feature of the decay amplitude
in Eq.~\eqref{Eq:DA_Bd2JpsiKS} is the suppression of the $a'e^{i\theta'}e^{i\gamma}$ term
by the tiny factor $\epsilon=0.0536 \pm 0.0003$. Consequently, $\phi_d$ can be extracted
with the help of Eq.~\eqref{AM-ideal} up to corrections of ${\cal O}(\epsilon a')$.

As was pointed out in Ref.~\cite{RF-psiK}, the decay \mbox{$B^0_s\to J/\psi K_{\rm S}^0$} is 
related to \mbox{$B^0_d\to J/\psi K_{\rm S}^0$} through the $U$-spin symmetry of strong 
interactions. It originates from \mbox{$\bar b\to \bar d c \bar c$} transitions and therefore has
a CKM structure which is different from \mbox{$B^0_d\to J/\psi K_{\rm S}^0$.} In analogy to 
Eq.~\eqref{Eq:DA_Bd2JpsiKS}, we write
\begin{equation}
A\left(B_s^0\rightarrow J/\psi K_{\mathrm S}^0\right) = - \lambda \mathcal{A}
\left[1- a e^{i\theta}e^{i\gamma}\right]\:,
\end{equation}
where the hadronic parameters are defined as their $B^0_d\to J/\psi K_{\rm S}^0$ counterparts. 
In contrast to Eq.~\eqref{Eq:DA_Bd2JpsiKS}, there is no $\epsilon$ factor present in front of 
the second term, thereby ``magnifying" the penguin effects. On the other hand, the $\lambda$
in front of the overall amplitude suppresses the branching ratio with respect to 
$B^0_d\to J/\psi K_{\rm S}^0$. 

The $U$-spin symmetry of strong interactions implies
\begin{equation}\label{a-rel}
a'e^{i\theta'}=ae^{i\theta}\:.
\end{equation}
In the factorisation approximation the hadronic form factors and decay constants cancel in 
the above amplitude ratios \cite{RF-psiK}, i.e.\ $U$-spin-breaking corrections enter 
$ae^{i\theta}$ through non-factorisable effects only. On the other hand, the relation
\begin{equation}\label{Aprime-A-rel}
\mathcal{A}'=\mathcal{A}
\end{equation}
is already in factorisation affected by $SU(3)$-breaking effects, 
entering through hadronic form factors as we will discuss in more detail below. 

It is well known that the factorisation approximation does not reproduce the branching ratios 
of $B\to J/\psi K$ decays well, thereby requiring large non-factorisable effects. Furthermore, 
the QCD penguin matrix elements of the current--current tree operators, which are usually assumed 
to yield the potential enhancement for the penguin contributions, vanish in naive factorisation 
for $B\to J/\psi K$ decays. Consequently, large non-factorisable contributions may also affect the
penguin parameters $a'e^{i\theta'}$ and $ae^{i\theta}$, thereby enhancing them from the smallish
values in factorisation, and Eq.~(\ref{a-rel}) may receive sizeable corrections -- despite the cancellation
of form factors and decay constants in factorisation. 

Making the replacements
\begin{equation}\label{replacement}
B^0_s\to J/\psi K_{\rm S}^0: \, b_f e^{i\rho_f} \, \rightarrow  \, a e^{i\theta}\:, \qquad
B^0_d\to J/\psi K_{\rm S}^0: \,  b_f e^{i\rho_f} \, \rightarrow \, - \epsilon a' e^{i\theta'}\:,
\end{equation}
we may apply the formalism introduced in Section~\ref{sec:hadr}, yielding the following phase 
shifts:
\begin{equation}\label{tan-phis}
\tan  \Delta\phi_s^{\psi K_{\rm S}^0}  =  \frac{-2 a\cos\theta\sin\gamma+a^2\sin2\gamma}{1-
2a\cos\theta\cos\gamma+a^2\cos2\gamma} =
-2a\cos\theta\sin\gamma-a^2\cos2\theta\sin 2\gamma +{\cal O}(a^3)\:,
\end{equation}
\begin{equation}\label{tan-phid}
\tan  \Delta\phi_d^{\psi K_{\rm S}^0}  = \frac{2 \epsilon a'\cos\theta'\sin\gamma+\epsilon^2 a'^2
\sin2\gamma}{1+2\epsilon a'\cos\theta'\cos\gamma+\epsilon^2 a'^2\cos2\gamma} 
=2\epsilon a'\cos\theta'\sin\gamma +{\cal O}(\epsilon^2 a'^2)\:.
\end{equation}
The expansions in terms of the penguin parameters show an interesting feature: the 
phase shifts are maximal for a strong phase difference around $0^{\circ}$ or $180^{\circ}$. 
Conversely, the penguin shifts will be tiny for values around $90^{\circ}$ or $270^\circ$, 
even for sizeable $a^{(')}$. The $\Delta\phi_s^{\psi K_{\rm S}^0}$ and 
$\Delta\phi_d^{\psi K_{\rm S}^0}$ enter 
\begin{equation}\label{phi-eff}
\phi_{s,\psi K_{\rm S}^0}^{\rm eff} = \phi_s+\Delta\phi_s^{\psi K_{\rm S}^0}\:, \qquad
\phi_{d,\psi K_{\rm S}^0}^{\rm eff} = \phi_d+\Delta\phi_d^{\psi K_{\rm S}^0}
\end{equation}
in the expressions corresponding to Eq.~\eqref{phiq-eff-def}. These  ``effective" mixing phases
are convenient for the presentation of the experimental results \cite{LHCb-psi-rho}.

\boldmath
\subsection{Branching Ratio Information}
\unboldmath
The $B^0_s\to J/\psi K_{\rm S}^0$ decay channel has been observed by the 
CDF \cite{Aaltonen:2011sy} and LHCb \cite{LHCb-BspsiK-lifetime} collaborations, 
and measurements of the time-integrated untagged rate
\cite{DFN}
\begin{equation}\label{time-int}
\mathcal{B}(B_s\to J/\psi K_{\rm S}^0) \equiv \frac{1}{2}\int_0^\infty 
\langle \Gamma(B_s(t)\to J/\psi K_{\rm S}^0) \rangle d t
\end{equation}
with
\begin{equation}
\langle \Gamma(B_s(t)\to J/\psi K_{\rm S}^0) \rangle \equiv 
\Gamma(B^0_s(t)\to J/\psi K_{\rm S}^0) +  \Gamma(\bar B^0_s(t)\to J/\psi K_{\rm S}^0) 
\end{equation}
were performed, resulting in the world average \cite{Agashe:2014kda}
\begin{equation}
\mathcal{B}(B_s\to J/\psi K_{\rm S}^0) = (1.87 \pm 0.17)\times 10^{-5}\:.
\end{equation}

Information on the penguin parameters is also encoded in this observable, thereby complementing
the CP asymmetries. In view of the sizeable decay width difference $\Delta\Gamma_s$ of the 
$B_s$-meson system, which is described by the parameter \cite{Amhis:2012bh}
\begin{equation}
y_s \equiv \frac{\Delta\Gamma_s}{2\Gamma_s} = 0.0608 \pm 0.0045\:,
\end{equation}
the ``experimental" branching ratio \eqref{time-int} has to be distinguished 
from the ``theoretical" branching ratio defined by the untagged decay rate at time $t=0$ 
\cite{RF-psiK}. The conversion of one branching ratio concept into the other can be done 
with the help of the following expression \cite{BR-paper}:
\begin{equation}\label{Eq:BR_correction}
\mathcal{B}\left(B_s\to J/\psi K_{\rm S}^0\right)_{\text{theo}} = 
\left[\frac{1-y_s^2}{1+\mathcal{A}_{\Delta\Gamma}(B_s\to J/\psi K_{\rm S}^0) \, y_s }\right]
\mathcal{B}\left(B_s\to J/\psi K_{\rm S}^0\right)\:.
\end{equation}
The observable $\mathcal{A}_{\Delta\Gamma}(B_s\to J/\psi K_{\rm S}^0)$ 
depends also on the penguin parameters, as can be seen in Eq.~(\ref{Eq:ADG}).

The effective lifetime 
\begin{align}
\tau_{J/\psi K_{\rm S}^0}^{\text{eff}} & \equiv  
\frac{\int_0^\infty t\,\langle \Gamma(B_s(t)\to J/\psi K_{\rm S}^0) \rangle\, dt}
	{\int_0^\infty \langle \Gamma(B_s(t)\to J/\psi K_{\rm S}^0) \rangle\, dt}\\
& = \frac{\tau_{B_s}}{1-y_s^2}\left[\frac{1+2\mathcal{A}_{\Delta\Gamma}(B_s\to J/\psi K_{\rm S}^0)
\:y_s + y_s^2}{1+\mathcal{A}_{\Delta\Gamma}(B_s\to J/\psi K_{\rm S}^0)\:y_s}\right]
\end{align}
allows us to determine $\mathcal{A}_{\Delta\Gamma}(B_s\to J/\psi K_{\rm S}^0)$, 
thereby fixing the conversion factor in Eq.~\eqref{Eq:BR_correction} \cite{BR-paper}.
The LHCb collaboration has performed the first measurement of this 
quantity \cite{LHCb-BspsiK-lifetime}:
\begin{equation}\label{Eq:tauEff_Bs2JpsiKs}
\tau_{J/\psi K_{\mathrm S}^0}^{\text{eff}} = \left(1.75 \pm 0.12 \pm 0.07\right)\,\text{ps}\:,
\end{equation}
corresponding to
\begin{equation}\label{ADG-exp}
\mathcal{A}_{\Delta\Gamma}(B_s\to J/\psi K_{\rm S}^0)= 2.1 \pm 1.6\:.
\end{equation}
In view of the large uncertainty of this measurement, we shall 
rely directly on Eq.~\eqref{Eq:ADG} with 
Eq.~\eqref{replacement} in the numerical analysis performed in Section~\ref{ssec:constr}. 

In order to utilise the branching ratio information, we construct the observable
\begin{equation}\label{Eq:Hobs_Def}
H  \equiv  \frac{1}{\epsilon} \left|\frac{\mathcal{A}'}{\mathcal{A}}\right|^2
\frac{\text{PhSp}\left(B_d\rightarrow J/\psi K_{\mathrm S}^0\right)}
{\text{PhSp}\left(B_s\rightarrow J/\psi K_{\mathrm S}^0\right)}
\frac{\tau_{B_d}}{\tau_{B_s}}
\frac{\mathcal{B}\left(B_s\rightarrow J/\psi K_{\mathrm S}^0\right)_{\text{theo}}}
{\mathcal{B}\left(B_d\rightarrow J/\psi K_{\mathrm S}^0\right)_{\text{theo}}}\:,
\end{equation}
where $\tau_{B_q}$ is the $B_q$ lifetime and $\text{PhSp}\left(B_q\rightarrow J/\psi X\right)$ 
denotes the phase-space function for these decays \cite{RF-psiK}.
In terms of the penguin parameters, we obtain
\begin{equation}\label{Eq:Hobs}
H =  \frac{1-2\:a\cos\theta\cos\gamma+a^2}{1+2\epsilon a'\cos\theta'\cos\gamma+\epsilon^2 
a^{\prime2}} = -\frac{1}{\epsilon}\frac{{\cal A}_{\rm CP}^{\rm dir}(B_d\to J/\psi 
K_{\rm S})}{{\cal A}_{\rm CP}^{\rm dir}(B_s\to J/\psi K_{\rm S})},
\end{equation}
where we also give the relation to the direct CP asymmetries of the decays at hand.
Keeping $a$ and $\theta$ as free parameters, the following lower bound arises 
\cite{Fleischer:1997um, Fleischer:2004rn}:
\begin{equation}\label{H-bound}
H \geq \frac{1+\epsilon^2+2 \epsilon \cos^2\gamma - (1+\epsilon) \sqrt{1-2 \epsilon +\epsilon^2+4 
\epsilon  \cos^2\gamma}}{2\epsilon ^2\left(1-\cos^2\gamma\right)}\:,
\end{equation}
which corresponds to $H \geq 0.872$ for $\gamma = 70^{\circ}$.

The determination of $H$ from the experimentally measured branching ratios
is affected by $U$-spin-breaking corrections which enter through the ratio 
$|\mathcal{A}'/\mathcal{A}|$. Consequently, $H$ is not a particularly clean observable.
On the other hand, the analysis of the direct and mixing-induced CP asymmetries does 
not require knowledge of $|\mathcal{A}'/\mathcal{A}|$.

\boldmath
\subsection{Determination of $\gamma$ and the Penguin Parameters}
\unboldmath
If we complement the ratio $H$ with the direct and mixing-induced CP asymmetries of the
$B^0_s\to J/\psi K_{\rm S}^0$ channel, we have sufficient information to determine
$\gamma$ and the penguin parameters $a$ and $\theta$ by means of the $U$-spin relation
in Eq.~\eqref{a-rel} \cite{RF-psiK}. In this strategy, $\phi_s$ serves as an input, where we
may either use its SM value in Eq.~\eqref{phis-SM} or the value extracted from
experimental data, as discussed in Section~\ref{sec:psiV}. We advocate the latter option
since it takes possible CP-violating NP contributions to $B^0_s$--$\bar B^0_s$ mixing into account. 

Although $\gamma$ can be extracted with this method at the LHCb upgrade, the corresponding
precision is not expected to be competitive with other strategies \cite{dBFK}. It is therefore
advantageous to employ $\gamma$ as an input. Using data from pure tree decays of the kind 
$B\to D^{(*)}K^{(*)}$, the following averages were obtained by the
CKMfitter and UTfit  collaborations:
\begin{equation}\label{gamma-range}
\gamma  = (70.0_{-9.0}^{+7.7})^{\circ}  \quad\text{(CKMfitter \cite{Charles:2011va})}\:, \qquad
\gamma  = (68.3 \pm 7.5)^{\circ} \quad\text{(UTfit \cite{Bevan:2014cya})\:.}
\end{equation}
For the numerical analysis in this paper, we shall use the CKMfitter result in view of the
larger uncertainty. By the time of the LHCb upgrade and Belle II era, much more precise 
measurements of $\gamma$ from pure tree decays will be available (see 
Section~\ref{ssec:scene}).

Once the direct and mixing-induced CP asymmetries of the $B^0_s\to J/\psi K_{\rm S}^0$ 
channel have been measured, Eqs.~\eqref{AD-expr} and \eqref{AM-expr} can be used 
with Eq.~\eqref{replacement} to determine $a$ and $\theta$ in a theoretically clean way. 
Employing the $U$-spin relation \eqref{a-rel} allows us to convert these parameters into 
the phase shift $\Delta\phi_d^{\psi K_{\rm S}^0}$, and thus to include the penguin effects in the 
determination of $\phi_d$.

\boldmath
\subsection{Constraining the Penguin Effects through Current Data}\label{ssec:constr}
\unboldmath
As a measurement of CP violation in $B^0_s\to J/\psi K_{\rm S}^0$ is not yet available, the 
$U$-spin strategy sketched above cannot yet be implemented in practice. However, in order to 
already obtain information on the size of the penguin parameters $a$ and $\theta$ and
their impact on high-precision studies of CP violation, we may use experimental data for 
decays which have dynamics similar to $B^0_s\to J/\psi K_{\rm S}^0$. 

\begin{figure}[tp]
\center
\includegraphics[width=0.30\textwidth]{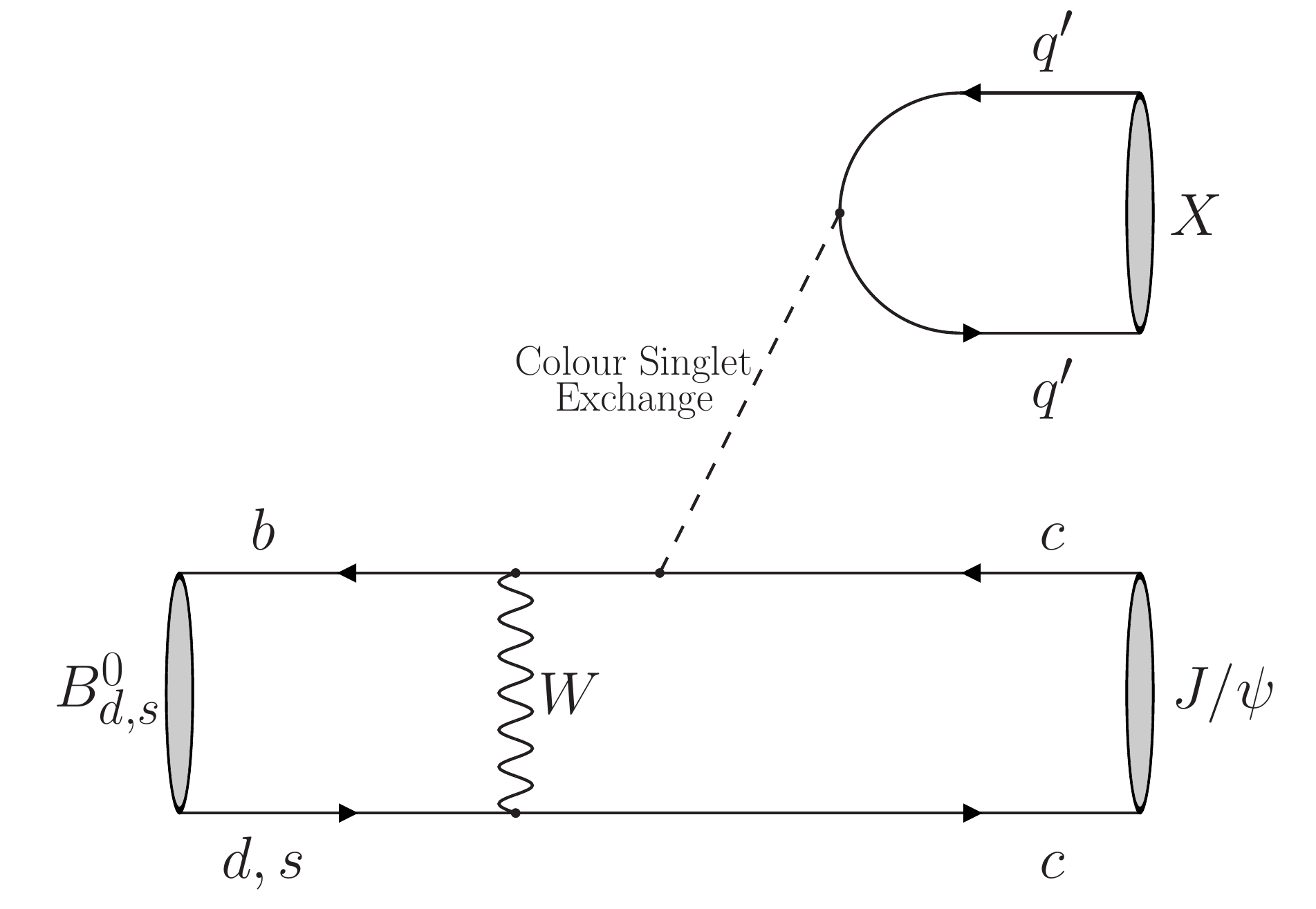}
\hfill
\includegraphics[width=0.30\textwidth]{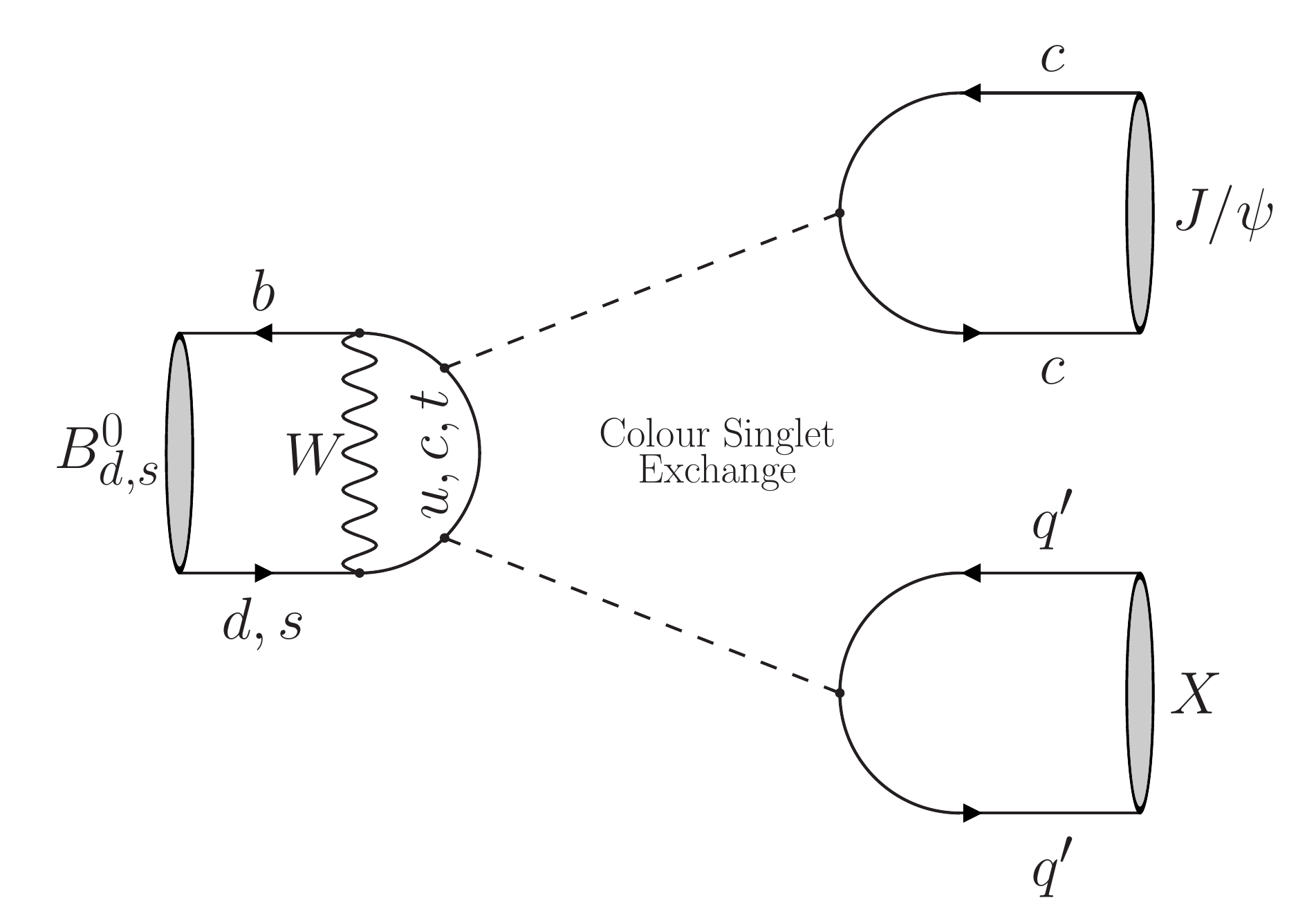}
\hfill
\includegraphics[width=0.30\textwidth]{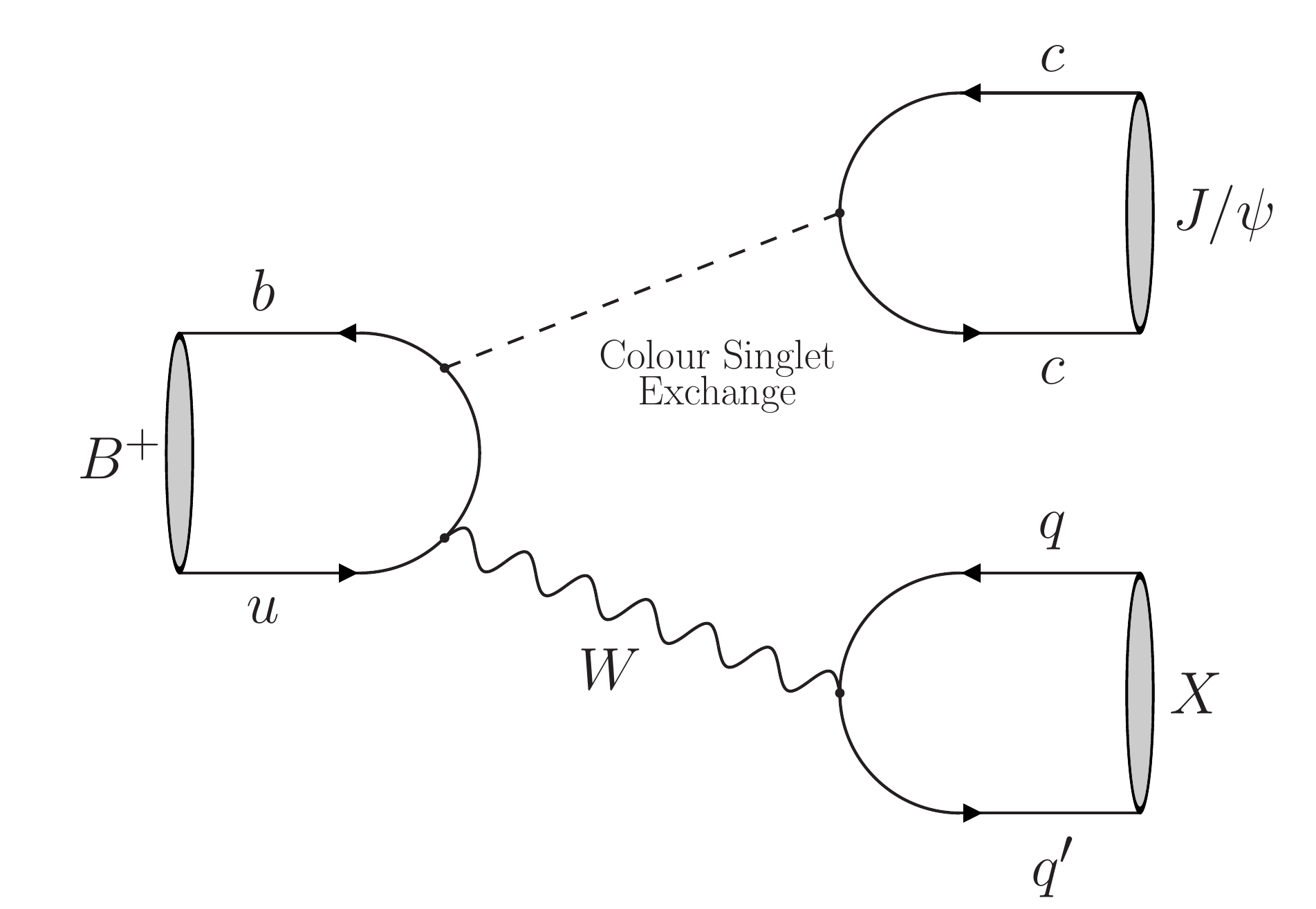}
\caption{Illustration of additional decay topologies contributing to some of the \mbox{$B\to J/\psi X$} channels: exchange (left), penguin annihilation (middle) and annihilation (right).}
\label{Fig:Feynman_Additional}
\end{figure}

If we replace the strange spectator quark with a down quark, as proposed in Ref.~\cite{RF-ang}, 
we obtain the $B^0_d\to J/\psi \pi^0$ decay \cite{CPS}, which is the vector--pseudo-scalar 
counterpart of the vector--vector mode $B^0_d\to J/\psi \rho^0$. 
The $B^0_d\to J/\psi \pi^0$ mode has contributions from penguin annihilation and 
exchange topologies, illustrated in Fig.~\ref{Fig:Feynman_Additional}, which have no 
counterpart in \mbox{$B^0_s\to J/\psi K_{\rm S}^0$} and are expected to be small. They 
can be probed through the $B^0_s\to J/\psi \pi^0$ 
decay (and $B^0_s\to J/\psi \rho^0$ for $B^0_d\to J/\psi \rho^0$) \cite{FFJM}. 
First measurements of CP violation in $B^0_d\to J/\psi \pi^0$ were reported by 
the BaBar and Belle collaborations:
\begin{equation}\label{jpsipi-meas}
\mathcal{A}_{\rm CP}^{\rm dir}(B_d\to J/\psi\pi^0) =
\begin{cases}
-0.08 \pm 0.16 \pm 0.05\: & \mbox{(Belle \cite{Lee:2007wd})}\\
-0.20 \pm 0.19 \pm 0.03\: & \mbox{(BaBar \cite{Aubert:2008bs})}
\end{cases}
\end{equation}
\begin{equation}\label{jpsipi-meas-mix}
\mathcal{A}_{\rm CP}^{\rm mix}(B_d\to J/\psi\pi^0) =
\begin{cases}
\phantom{-}0.65 \pm 0.21 \pm 0.05\: & \mbox{(Belle \cite{Lee:2007wd})}\\
\phantom{-}1.23 \pm 0.21 \pm 0.04\: & \mbox{(BaBar \cite{Aubert:2008bs})}\:.
\end{cases}
\end{equation}
The results for the mixing-induced CP asymmetry are not in good agreement
with each other, with the BaBar result lying outside the physical region. 
The Heavy Flavour Averaging Group (HFAG) has refrained from inflating the uncertainties in
their average, giving $\mathcal{A}_{\rm CP}^{\rm mix}(B_d\to J/\psi\pi^0) = 0.93 \pm 0.15$
\cite{Amhis:2012bh}. The Belle II experiment will hopefully clarify this unsatisfactory situation.

The charged counterpart $B^+\to J/\psi \pi^+$ of $B^0_d\to J/\psi \pi^0$ also has
dynamics similar to $B^0_s\to J/\psi K_{\rm S}^0$ but --- as it is the decay of a charged $B$ 
meson --- does not exhibit mixing-induced CP violation. It receives additional contributions 
from an annihilation topology, illustrated in Fig.~\ref{Fig:Feynman_Additional}, which 
arises with the same CKM factor $V_{ud}^{\phantom{*}}V_{ub}^*$ as the penguin topologies with internal 
up-quark exchanges, contributing similarly to the penguin parameter $a_{\rm c}e^{i\theta_{\rm c}}$
(defined in analogy to Eq.~(\ref{Eq:Penguin_Def})). If this parameter is determined from the
charged $B^+\to J/\psi \pi^+$, $B^+\to J/\psi K^+$ decays and compared with the other
penguin parameters, footprints of the annihilation topology could be detected. 
In view of the present uncertainties, we neglect the annihilation topology, like the
contributions from the exchange  and penguin annihilation topologies in $B^0_d\to J/\psi \pi^0$.
In Appendix~\ref{App:annihil}, we give a more detailed discussion of the annihilation 
contribution and its importance based on constraints from current
data, which do not indicate any enhancement.

We shall also add data for the $B^+\to J/\psi K^+$ (neglecting again the corresponding
annihilation contribution) and $B^0_d\to J/\psi K^0$ modes to the global analysis, 
although the penguin contributions are doubly Cabibbo-suppressed in these decays. 

Using the $SU(3)$ flavour symmetry and assuming both vanishing non-factorisable corrections
and vanishing exchange and annihilation topologies, the decays listed above are characterised 
by a universal set of penguin parameters 
$(a,\theta)$, which can be extracted from the input data through a global $\chi^2$ fit. 
The resulting picture extends and updates the previous analyses of Refs.~\cite{CPS, FFJM}.

\begin{figure}
\center
\includegraphics[width=0.75\textwidth]{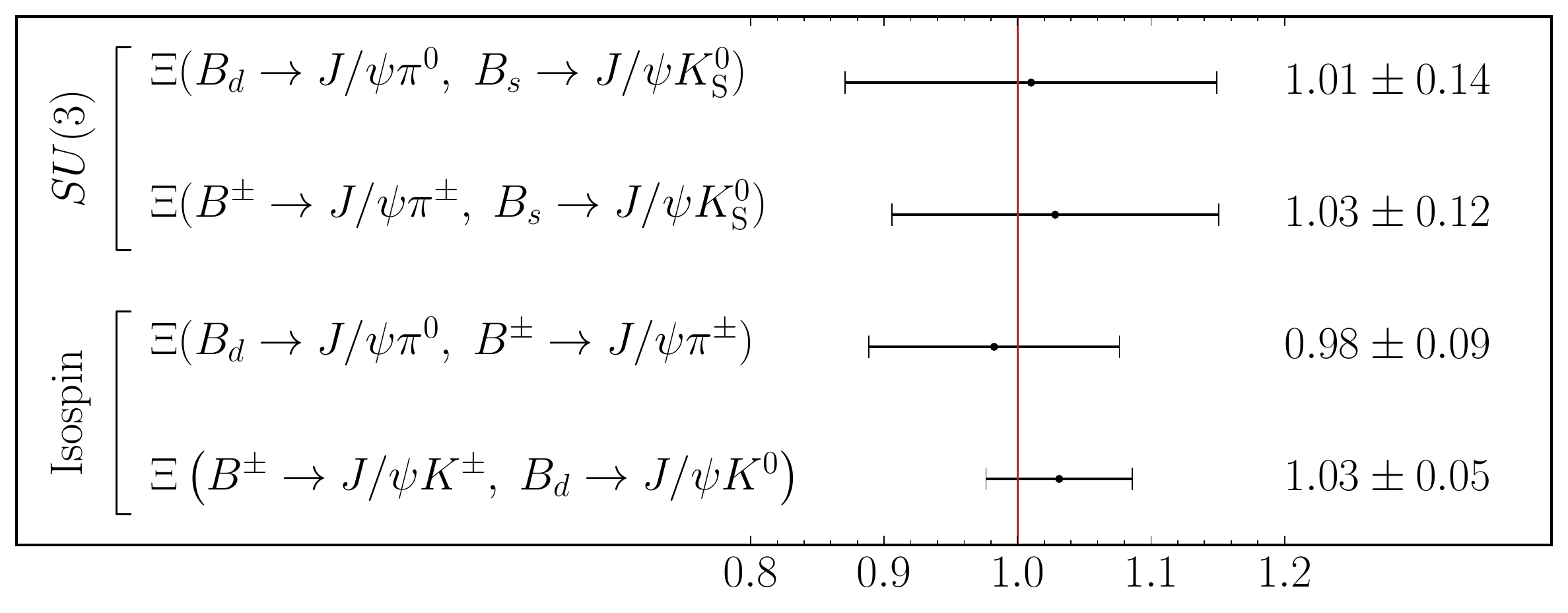}
\caption{Overview of the different ratios defined in Eq.~\eqref{Eq:SU3_RatioTest}.
In the limit where we neglect the contributions from additional decay topologies and 
assume perfect flavour symmetry for the spectator quarks, the ratios equal unity.}
\label{Fig:SU3_comparison}
\end{figure}

\begin{figure}
\center
\includegraphics[width=0.75\textwidth]{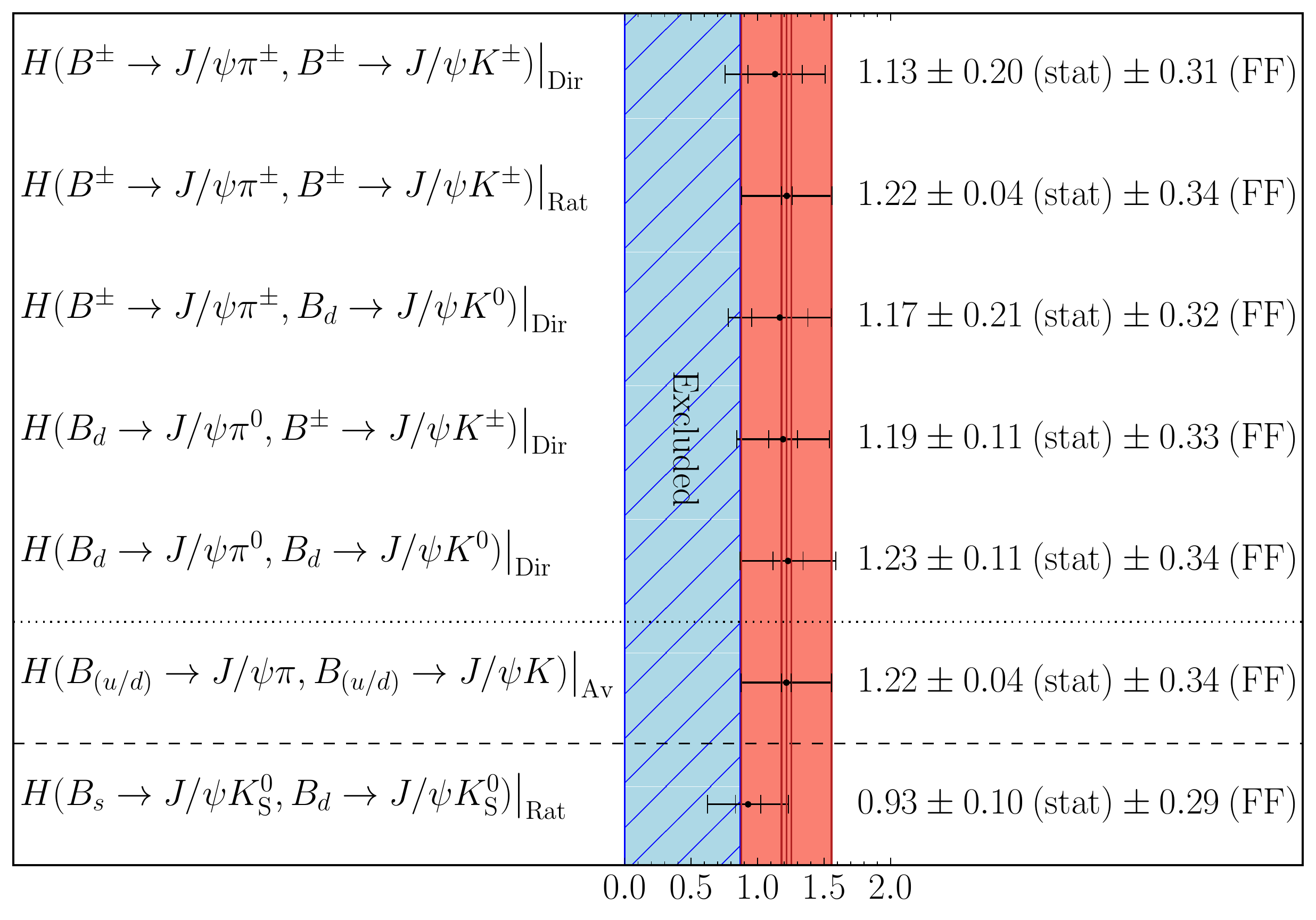}
\caption{$H$ observables which can be constructed from the available 
branching ratio information for \mbox{$B_q\to J/\psi P$} modes.
The label ``Dir" indicates that $H$ is determined from direct branching 
fraction measurements, whereas the label ``Rat" is used 
for $H$ observables calculated from a ratio of branching fractions.
The inner uncertainty bars indicate the statistical uncertainty whereas the outer ones give the 
total uncertainty, including the common uncertainty due to the form factors.
The red band indicates the average $H$ observable of the 
\mbox{$B_{(u/d)}\rightarrow J/\psi(\pi/K)$} modes. The hatched blue region is 
excluded by Eq.~\eqref{H-bound}.}
\label{Fig:Hobs}
\end{figure}

A first consistency check is provided by the ratios
\begin{equation}\label{Eq:SU3_RatioTest}
\Xi(B_q\rightarrow J/\psi X, B_{q'}\rightarrow J/\psi Y)\equiv
\frac{\text{PhSp}\left(B_{q'}\rightarrow J/\psi Y\right)}{\text{PhSp}\left(B_q\rightarrow J/\psi X\right)}
\frac{\tau_{B_{q'}}}{\tau_{B_q}}
\frac{\mathcal{B}\left(B_q\rightarrow J/\psi X\right)_{\text{theo}}}
{\mathcal{B}\left(B_{q'}\rightarrow J/\psi Y\right)_{\text{theo}}}\:,
\end{equation}
involving decays which originate from the same quark-level processes but differ through 
their spectator quarks \cite{dBFK}. Neglecting exchange and annihilation 
topologies and assuming perfect flavour symmetry of strong interactions, these ratios equal one.
Within the uncertainties, this picture is supported by the data, as shown 
in Fig.~\ref{Fig:SU3_comparison}. In this compilation, the $B$-factory branching 
ratio measurements are corrected for the measured pair production asymmetry between 
$B_d^0\bar{B}_d^0$ and $B^+B^-$ \cite{Agashe:2014kda} at the $\Upsilon(4S)$ resonance. 
Note that the branching ratios for decays into final states with $K_{\rm S}^0$ or $\pi^0$ 
mesons have to be multiplied by a factor of two in Eq.~\eqref{Eq:SU3_RatioTest} to take 
the $K_{\rm S}^0$ and $\pi^0$ wave functions into account.

Let us now probe the penguin parameters through the various branching ratios. To this
end, we use ratios defined in analogy to $H$ in Eq.~\eqref{Eq:Hobs_Def}. The extraction
of these quantities from the data requires knowledge of the amplitude ratio 
$|\mathcal{A}'/\mathcal{A}|$, which is given in factorisation as follows \cite{RF-psiK}:
\begin{equation}\label{A-fact}
\left|\frac{\mathcal{A}'(B_{q'}\to J/\psi X)}{\mathcal{A}(B_q\to J/\psi Y)}\right|_{\rm fact} 
= \frac{f^+_{B_{q'}\rightarrow X}(m_{J/\psi}^2)}{f^+_{B_q\rightarrow Y}(m_{J/\psi}^2)}\:.
\end{equation}
The corresponding form factors have been calculated in the literature using a variety of techniques.
For our analysis, we take the results from light cone QCD sum rules (LCSR), 
which are typically calculated at $q^2=0$.
The relevant form factors are \mbox{$f_{B\to\pi}^+ (0)= 0.252_{-0.028}^{+0.019}$} 
\cite{Bharucha:2012wy}, \mbox{$f_{B\to K}^+ (0)= 0.34_{-0.02}^{+0.05}$} \cite{Khodjamirian:2010vf}
and \mbox{$f_{B_s\to K}^+(0) = 0.30_{-0.03}^{+0.04}$} \cite{Duplancic:2008tk}, where the first two 
describe transitions for both the $B_d^0$ and the $B^+$ mesons. The $q^2$ dependence 
of these form factors is parametrised by means of the BGL 
method described in Ref.~\cite{Ball:2006jz}.

Using these form factors and neglecting non-factorisable $SU(3)$-breaking effects, 
we obtain the various $H$ observables compiled in Fig.~\ref{Fig:Hobs}. With exception 
of the last entry, all $H$ observables share the same ratio $f_{B\to K}^+/f_{B\to\pi}^+$. 
Consequently, their central values and uncertainties are highly correlated.
However, even restricting the comparison to the statistical uncertainties shows an excellent
compatibility between the various $H$ results. The corresponding ratios are related to each
other through the isospin symmetry (neglecting additional topologies), and we obtain 
a consistent experimental picture. The
agreement with the last entry, which involves the decay $B^0_s\to J/\psi K^0_{\rm S}$
instead of the $B\to J/\psi \pi$ modes, suggests that non-factorisable $SU(3)$-breaking 
effects and the impact of additional decay topologies 
are small, thereby complementing the picture of Fig.~\ref{Fig:SU3_comparison}. 
The uncertainties are still too large to draw definite conclusions. 
\newline

For the global $\chi^2$ fit to extract the penguin parameters $a$ and $\theta$ we use the 
input quantities summarised in Table~\ref{Tab:Inputs_CP_B2PV}, and add the CKMfitter
result for $\gamma$ in Eq.~\eqref{gamma-range} as an asymmetric Gaussian 
constraint. As far as the $H$ 
observables are concerned, we employ the average of the 
$B_{(u/d)}\rightarrow J/\psi(\pi/K)$ combinations, which involve the same set of form 
factors (see Fig.~\ref{Fig:Hobs}), and the $H$ observable of the 
$B_{s,d}^0\rightarrow J/\psi K_{\rm S}^0$ system. The branching ratios entering 
the $H$ observables are complemented by the corresponding direct CP asymmetries. 

In order to add the mixing-induced CP asymmetry of the $B^0_d\to J/\psi \pi^0$ channel
to the fit, the $B^0_d$--$\bar B^0_d$ mixing phase $\phi_d$ is needed as an input. However, 
the measured CP-violating asymmetries of the $B^0_d\to J/\psi K_{\rm S}^0$ decay 
allow us to determine only the effective mixing phase\footnote{The numerical value in 
Eq.~\eqref{phid-eff} actually corresponds to the mixing-induced CP asymmetry 
$\mathcal{A}_{\rm CP}^{\rm mix}(B_d\rightarrow J/\psi K^0)$, which is an average
of $B^0_d\to J/\psi K_{\rm S}^0$ and $B^0_d\to J/\psi K_{\rm L}^0$ data \cite{Amhis:2012bh}.}
\begin{equation}\label{phid-eff}
\phi_{d,\psi K_{\rm S}^0}^{\rm eff}= \phi_d + \Delta\phi_d^{\psi K_{\rm S}^0} =(42.1\pm1.6)^\circ
\end{equation}
from  Eq.~\eqref{phiq-eff-def}. But --- if we express the phase shift 
$\Delta\phi_d^{\psi K_{\rm S}^0}$ in terms of the penguin parameters --- we may 
add this observable to our analysis.

\begin{table}[tp]
\center
\begin{tabular}{lcr}
\toprule
Observable & \multicolumn{2}{c}{Experimental result}\\
\midrule
\midrule
$\mathcal{A}_{\rm CP}^{\rm dir}(B^{\pm}\rightarrow J/\psi\pi^{\pm})$ & $-0.001 \pm 0.023$ & 
\cite{Agashe:2014kda}\\
$\mathcal{A}_{\rm CP}^{\rm dir}(B_d\rightarrow J/\psi\pi^0)$ & $-0.13 \pm 0.13$ 
& \cite{Agashe:2014kda}\\
$\mathcal{A}_{\rm CP}^{\rm mix}(B_d\rightarrow J/\psi\pi^0)$ & $\phantom{-}0.94 \pm 0.15$ & 
\cite{Agashe:2014kda}\\
\midrule
$\mathcal{A}_{\rm CP}^{\rm dir}(B^{\pm}\rightarrow J/\psi K^{\pm})$ & $-0.0030 \pm 0.0033$ 
& \cite{Agashe:2014kda}\\
$\mathcal{A}_{\rm CP}^{\rm dir}(B_d\rightarrow J/\psi K^0)$ & $\phantom{-}0.007 \pm 0.020$ & 
\cite{Amhis:2012bh}\\
$\mathcal{A}_{\rm CP}^{\rm mix}(B_d\rightarrow J/\psi K^0)$ & $-0.670 \pm 0.021$ & 
\cite{Amhis:2012bh}\\
\midrule
$H(B_{(u/d)}\rightarrow J/\psi(\pi/K))$ & $\phantom{-}1.22 \pm 0.34$ & Fig.~\ref{Fig:Hobs}\\
$H(B_{(s/d)}\rightarrow J/\psi K_{\rm S}^0)$ & $\phantom{-}0.93 \pm 0.31$ & Fig.~\ref{Fig:Hobs} \\
\bottomrule
\end{tabular}
\caption{Input quantities for the global $\chi^2$ fit to the penguin parameters $a$, $\theta$ 
and $\phi_d$.}
\label{Tab:Inputs_CP_B2PV}
\end{table}

\begin{figure}[tp]
\center
\includegraphics[width=0.75\textwidth]{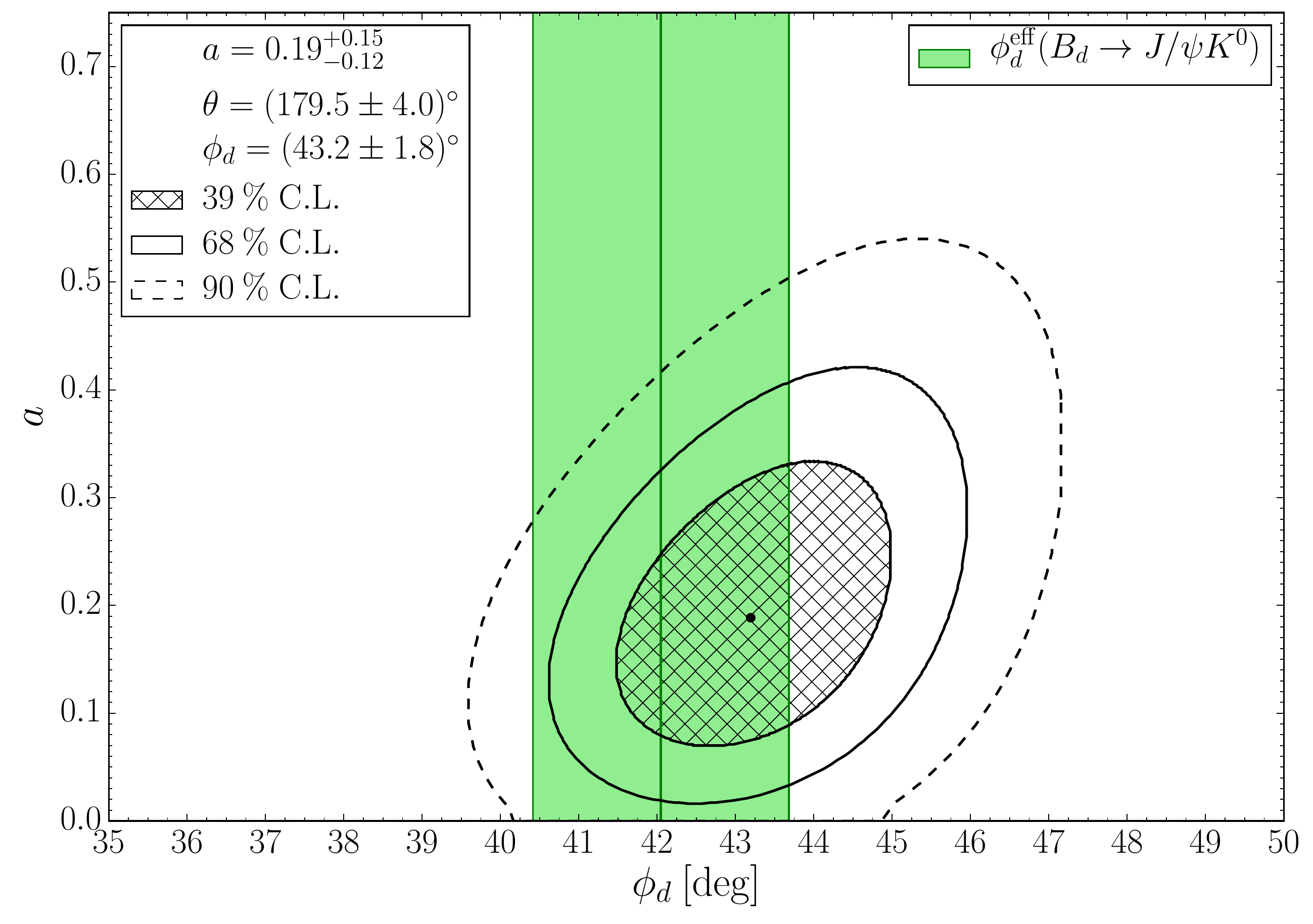}
\caption{Correlation between the $B^0_d$--$\bar B^0_d$ mixing phase $\phi_d$ and
the penguin parameter $a$ arising from the $\chi^2$ fit to current data as described in
the text.}
\label{fig:phid-a}
\end{figure}

\begin{figure}[tp]
\center
\includegraphics[width=0.75\textwidth]{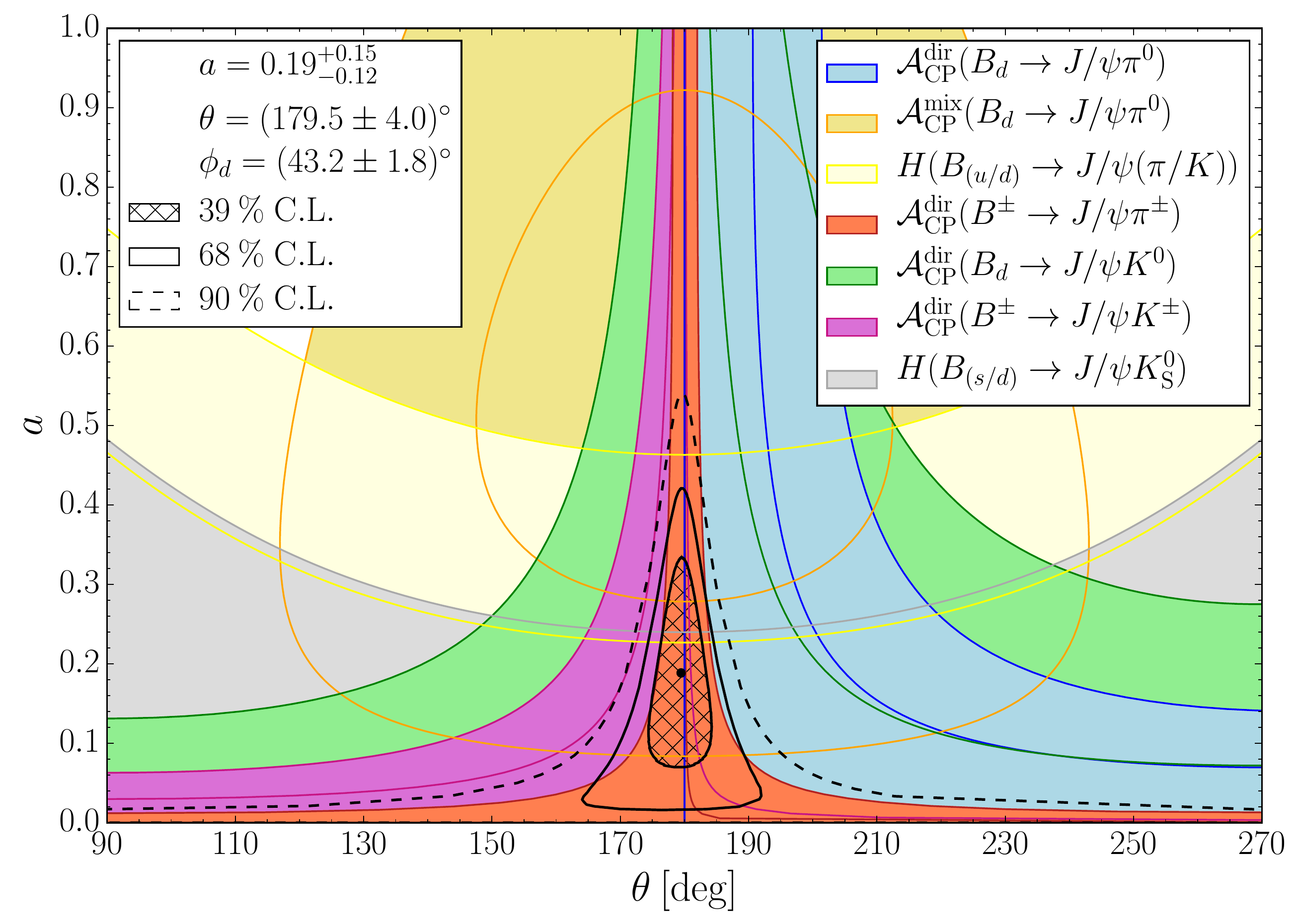}
\caption{Determination of the penguin parameters $a$ and $\theta$ through intersecting contours 
derived from CP asymmetries and branching ratios of $B_q\to J/\psi P$ decays. We show
also the confidence level contours obtained from a $\chi^2$ fit to the data.
To improve the visualisation, the allowed range for $a$ has been extended to 1.}
\label{Fig:B2PV}
\end{figure}

\begin{figure}[tp]
\center
\includegraphics[width=0.75\textwidth]{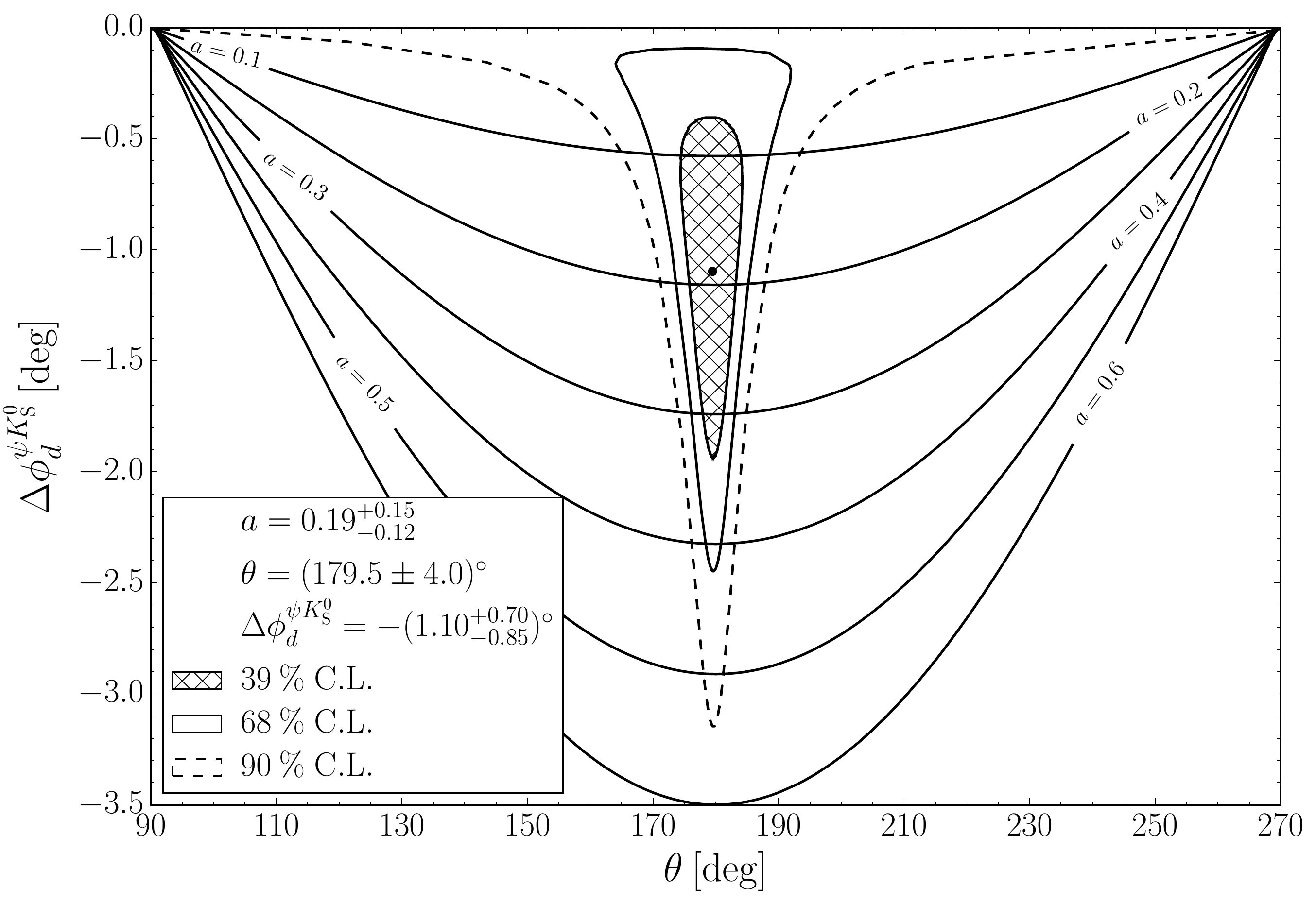}
\caption{Constraints on  $\Delta\phi_d^{\psi K_{\mathrm S}^0}$ as a function of 
the strong phase $\theta$ arising from the $\chi^2$ fit to the data.
Superimposed are the contour levels for the penguin parameter $a$.}\label{Fig:B2PV_Deltaphi}
\end{figure}

The global fit yields $\chi^2_{\text{min}} = 2.6$ for four degrees of freedom $(a,\theta, \phi_d,\gamma)$, 
indicating good agreement between the different input quantities. It results in the solutions
\begin{equation}\label{Eq:B2PV_chi2fit}
a = 0.19^{+0.15}_{-0.12} \:,\qquad \theta = \left(179.5 \pm 4.0\right)^{\circ}\:
\end{equation}
and
\begin{equation}\label{Eq:phid_chi2fit}
\phi_d = \left(43.2^{+1.8}_{-1.7}\right)^{\circ}\:,
\end{equation}
while $\gamma$ is constrained to the input in Eq.~\eqref{gamma-range}.
In Fig.~\ref{fig:phid-a}, we show the correlation between $\phi_d$ and $a$. The value of 
$\phi_d$ in Eq.~\eqref{Eq:phid_chi2fit} will serve as an input in Section~\ref{sec:psiV}.
Following Ref.~\cite{FFJM}, we illustrate the various constraints entering the fit 
through contour bands of the individual observables in Fig.~\ref{Fig:B2PV}. For the 
${\cal A}_{\rm CP}^{\rm mix}(B_d\to J/\psi \pi^0)$ range, we have used the value of 
$\phi_d$ in Eq.~\eqref{Eq:phid_chi2fit}. In comparison with the analysis of Ref.~\cite{FFJM}, 
the penguin parameters are now constrained in a more stringent way. 
The penguin parameters in Eq.~\eqref{Eq:B2PV_chi2fit} result in the following 
penguin phase shift:
\begin{equation}\label{Dphi-fit}
\Delta\phi_d^{\psi K_{\rm S}^0} = -\left(1.10^{+0.70}_{-0.85}\right)^{\circ}\:,
\end{equation}
with confidence level contours shown in Fig.~\ref{Fig:B2PV_Deltaphi}.

\begin{figure}[tp]
\center
\includegraphics[width=0.75\textwidth]{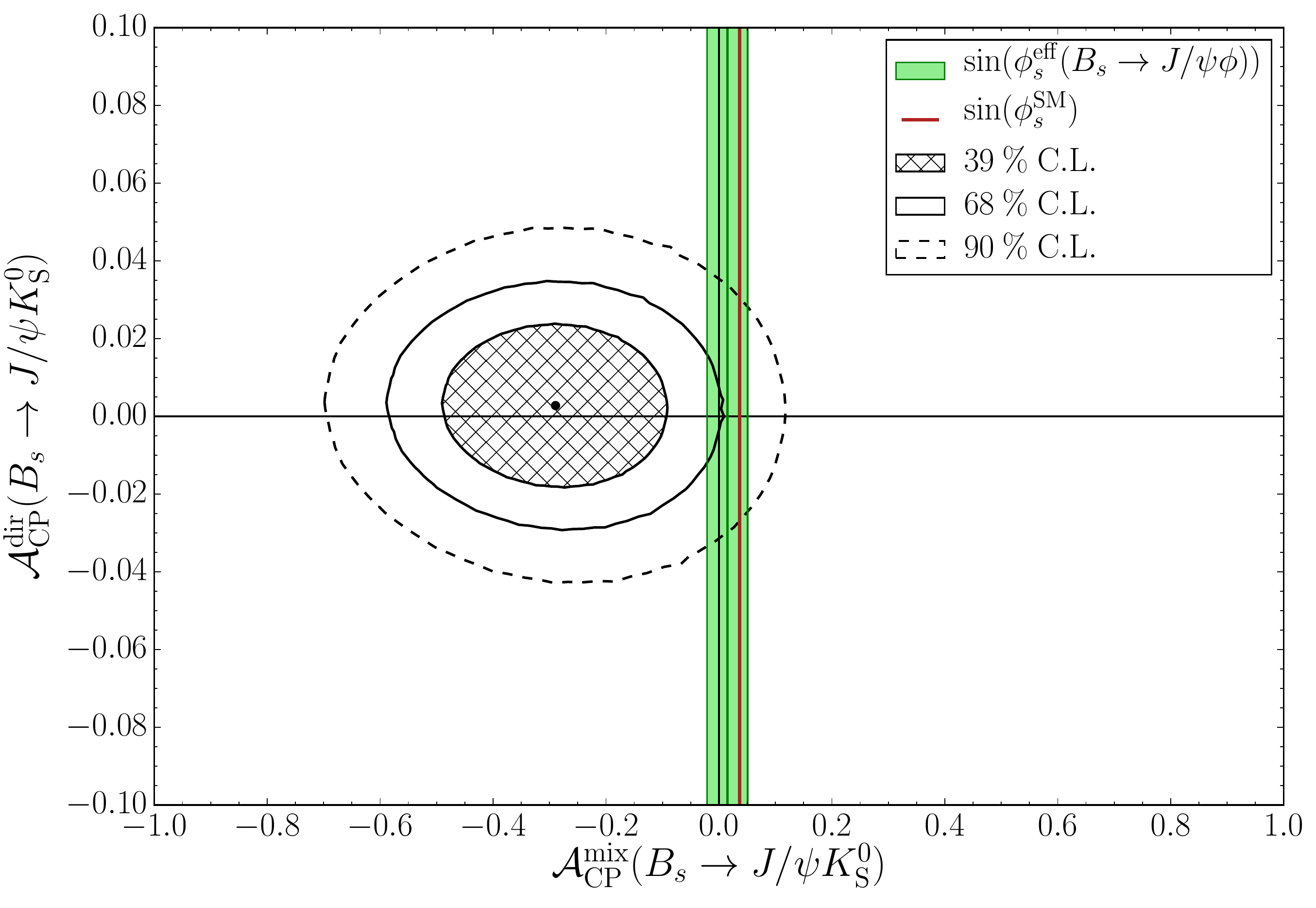}
\caption{Prediction of CP violation in
\mbox{$B_s^0\to J/\psi K^0_{\mathrm S}$} following from the global
$\chi^2$ fit to the present data as discussed in Section~\ref{ssec:constr}.}\label{Fig:Prediction}
\end{figure}

%
%
%
\boldmath
\subsection{Benchmark Scenario for $B^0_{d,s}\to J/\psi K_{\rm S}^0$}\label{ssec:scene}
\unboldmath
Let us conclude the analysis of the penguin effects in $B^0_d\to J/\psi K_{\rm S}^0$ by 
discussing a future benchmark scenario pointing to the LHCb upgrade era. Using the 
results in Eq.~\eqref{Eq:B2PV_chi2fit} and assuming 
the SM value for $\phi_s$ in Eq.~\eqref{phis-SM}, we obtain the following predictions:
\begin{alignat}{2}
\mathcal{A}_{\Delta\Gamma}(B_s\to J/\psi K_{\mathrm S}^0) = &\:& 0.957 & \pm 
0.061\:,\\
{\cal A}_{\rm CP}^{\rm dir}(B_s\to J/\psi K_{\mathrm S}^0) = &\:& 0.003 & 
\pm 0.021\:,\label{AD-pred}\\
{\cal A}_{\rm CP}^{\rm mix}(B_s\to J/\psi K_{\mathrm S}^0) = &\:& -0.29\phantom{0} 
& \pm 0.20\:.\label{AM-pred}
\end{alignat}
The associated confidence level contours for 
${\cal A}_{\rm CP}^{\rm dir}(B_s\to J/\psi K_{\mathrm S}^0)$ and 
${\cal A}_{\rm CP}^{\rm mix}(B_s\to J/\psi K_{\mathrm S}^0)$ are 
shown in Fig.~\ref{Fig:Prediction}. 
Moreover, the penguin parameters in Eq.~\eqref{Eq:B2PV_chi2fit} yield 
\begin{equation}
\tau_{J/\psi K_{\mathrm S}^0}^{\text{eff}} = (1.603 \pm 0.010)~{\rm ps}\:,
\end{equation}
in agreement with the experimental result in Eq.~\eqref{Eq:tauEff_Bs2JpsiKs}.

In order to illustrate the potential of the $B^0_s\to J/\psi K_{\rm S}^0$ decay to extract the 
penguin parameters at the LHCb upgrade, let us assume that $\gamma$ has been
determined in a clean way from pure tree decays $B\to D^{(*)}K^{(*)}$ as
\begin{equation}\label{gamma-LHCbupgrade}
\gamma=(70\pm1)^\circ\:,
\end{equation}
and that the $B^0_s$--$\bar B^0_s$ mixing phase has been extracted from 
the $B^0_s\to J/\psi \phi$ angular analysis and the application of the strategies 
discussed in Sections~\ref{sec:psiV} and \ref{sec:road} to control the penguin effects as
\begin{equation}\label{phis-scen}
\phi_s=-\left(2.1\pm0.5|_{\rm exp}\pm 0.3 |_{\rm theo}\right)^\circ=-(2.1\pm0.6)^\circ\:.
\end{equation}
The experimental uncertainty projections for the LHCb upgrade are discussed in 
Ref.~\cite{LHCb-implications}. We consider our assessment of the theoretical uncertainty
of $\phi_s$ in Eq.~\eqref{phis-scen} as conservative. 

Let us assume that the CP-violating asymmetries of the 
$B^0_s\to J/\psi K_{\rm S}^0$ channel have been measured as follows:
\begin{equation}\label{CP-bench}
{\cal A}_{\rm CP}^{\rm dir}(B_s\to J/\psi K_{\mathrm S}^0) =  0.00 \pm 0.05\:, \qquad
{\cal A}_{\rm CP}^{\rm mix}(B_s\to J/\psi K_{\mathrm S}^0) =  -0.28 \pm 0.05\:,
\end{equation}
i.e.\ with the central values of Eqs.~\eqref{AD-pred} and \eqref{AM-pred}. In order to 
estimate the uncertainties, current LHCb measurements of CP violation in 
$B_s^0\to D_s^{\mp} K^{\pm}$  modes \cite{Aaij:2014fba} have been extrapolated
to the LHCb upgrade era, correcting for the $B_s^0\to J/\psi K_{\mathrm S}^0$ event yield 
 \cite{Aaij:2012nh}.

\begin{figure}[tp]
\center
\includegraphics[width=0.75\textwidth]{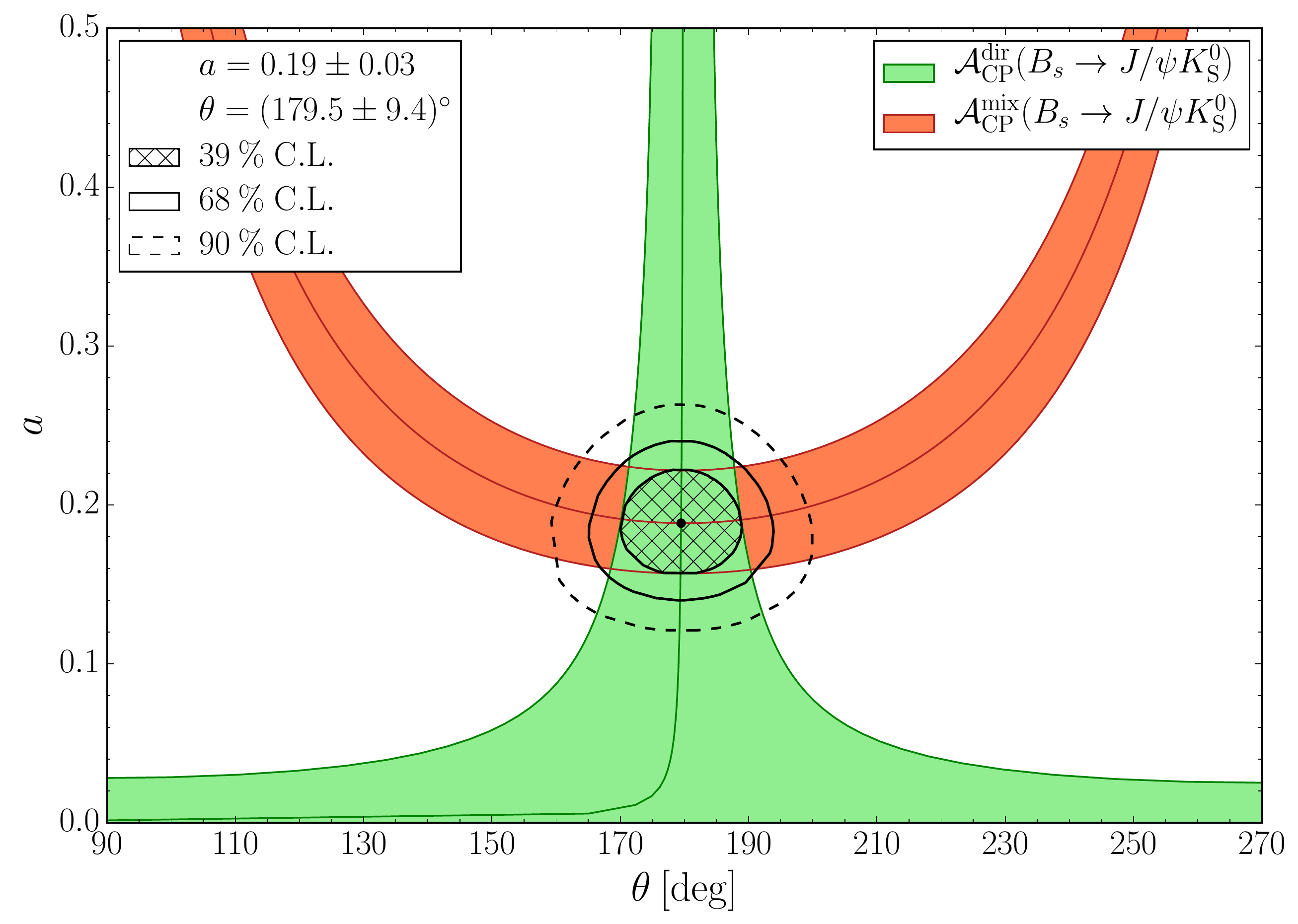}
\caption{Benchmark scenario illustrating the determination of the penguin parameters 
$a$ and $\theta$ from the CP asymmetries of the \mbox{$B_s^0\to J/\psi K^0_{\mathrm S}$} decay.}
\label{Fig:Bench_Bs}
\end{figure}

A $\chi^2$ fit to these observables 
would then yield
\begin{equation}\label{a-theta-bench}
a = 0.189^{+0.034}_{-0.032} \:,\qquad \theta = \left(179.5\pm 9.4\right)^{\circ}\:.
\end{equation}
The corresponding confidence level contours are shown in Fig.~\ref{Fig:Bench_Bs}.

\begin{figure}[tp]
\center
\includegraphics[width=0.75\textwidth]{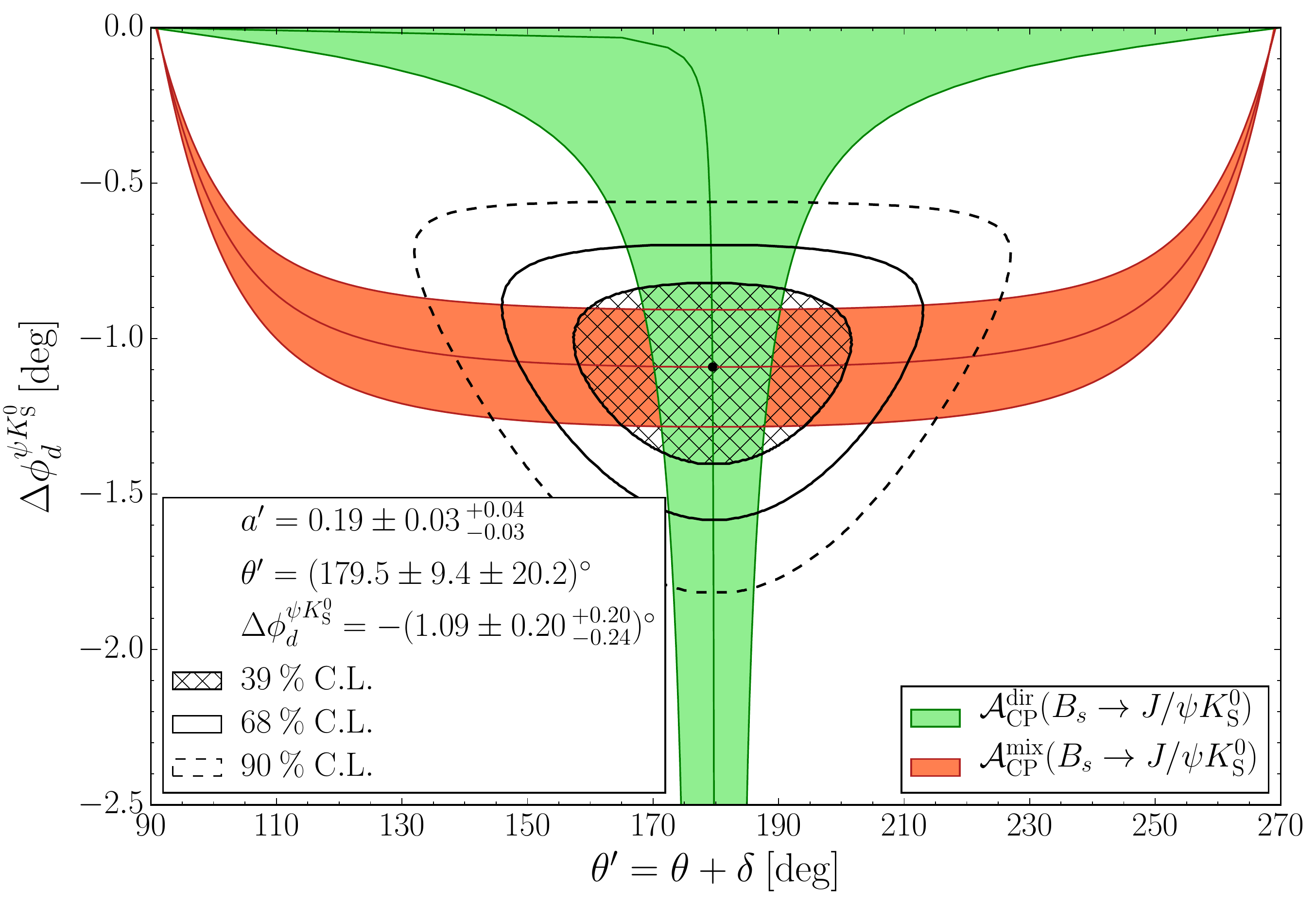}
\caption{Benchmark scenario illustrating the determination of $\Delta\phi_d^{\psi K_{\rm S}^0}$ 
from the CP asymmetries of the \mbox{$B_s^0\to J/\psi K_{\rm S}^0$} decay. The 
confidence level contours assume a 20\% uncertainty for $U$-spin breaking effects, 
parametrised through Eq.~\eqref{a-rel-break}.}
\label{Fig:Bench_Bd}
\end{figure}

In contrast to the fit in Fig.~\ref{Fig:B2PV}, this ``future" determination of $a$ and $\theta$ is 
theoretically clean. Using the $U$-spin relation \eqref{a-rel}, these parameters can be
converted into the penguin phase shift of $B^0_d\to J/\psi K_{\rm S}^0$. It is only at this point that 
potential $U$-spin-breaking effects enter. They can be included by introducing 
parameters $\xi$ and $\delta$ as follows:
\begin{equation}\label{a-rel-break}
a' = \xi\cdot a\:,\qquad\theta' = \theta + \delta\:.
\end{equation}
Assuming $\xi = 1.00 \pm 0.20$ and $\delta = (0 \pm 20)^{\circ}$, the results for $a$ and $\theta$ 
in Fig.~\ref{Fig:Bench_Bs} yield
\begin{equation}
\Delta\phi_d^{\psi K_{\rm S}^0} = -\left[1.09 \pm 0.20\:(\text{stat})_{-0.24}^{+0.20}\:(\text{U-spin})
\right]^{\circ},
\end{equation}
with the corresponding contours shown in Fig.~\ref{Fig:Bench_Bd}. In this benchmark 
scenario, the experimental and theoretical uncertainties are of the same size, with a 
total uncertainty of $0.3^\circ$ if added in quadrature. 
By the time such measurements will be available, we should have better experimental insights
into $U$-spin-breaking effects. As we will see in Section~\ref{Sec:B2JpsiRho}, 
already the currently available data for $B^0_d\to J/\psi\rho^0$ decays do not 
favour large effects in the $B^0_d\to J/\psi\rho^0$, $B^0_s\to J/\psi \phi$ system.

It is important to emphasise that the observable $H$ is {\it not} required in this analysis. 
Assuming Eq.~\eqref{a-rel}, it can rather
be determined with the help of Eq.~\eqref{Eq:Hobs}. As $a'$ enters there in combination with 
the tiny $\epsilon$ factor, the $U$-spin-breaking corrections have a negligible effect in this 
case. Using Eq.~\eqref{a-theta-bench}, we obtain 
\begin{equation}
H_{(a,\theta)}=1.172\pm0.037\:(a,\theta)\pm0.0016\:(\xi,\delta)\:.
\end{equation}
The comparison with Eq.~\eqref{Eq:Hobs_Def} then allows us to extract the ratio
\begin{equation}\label{A-det}
\left|\frac{\mathcal{A}'}{\mathcal{A}}\right|=
\sqrt{\epsilon \, H_{(a,\theta)} \frac{\text{PhSp}\left(B_s\rightarrow J/\psi 
K_{\mathrm S}^0\right)}{\text{PhSp}\left(B_d\rightarrow J/\psi K_{\mathrm S}^0\right)}
\frac{\tau_{B_s}}{\tau_{B_d}}\frac{\mathcal{B}\left(B_d\rightarrow J/\psi 
K_{\mathrm S}^0\right)_{\text{theo}}}{\mathcal{B}\left(B_s\rightarrow 
J/\psi K_{\mathrm S}^0\right)_{\text{theo}}}}\:.
\end{equation}
In order to illustrate the corresponding future experimental precision, 
we use the central values of the penguin parameters in Eq.~\eqref{a-theta-bench} and 
combine them with information on the ratio of branching fractions. 
The systematic uncertainty of all $B_s$ branching ratio measurements
is limited by the ratio $f_s/f_d = 0.259 \pm 0.015$ \cite{LHCb:2013lka} 
of fragmentation functions, which is required for normalisation purposes \cite{FST}.
At the LHCb upgrade, the experimental precision of the ratio of branching fractions entering
Eq.~\eqref{A-det} will be governed by that of $f_s/f_d$. Assuming no further improvement 
in the determination of this parameter, which is conservative, would result 
in the measurement
\begin{equation}
\left|\frac{\mathcal{A}'}{\mathcal{A}}\right|_{\rm exp}=1.160\pm0.035\:.
\end{equation}
The experimental uncertainty is about five times smaller than the current theoretical uncertainty
of the factorisation result
\begin{equation}\label{A-rat-fact-est}
\left|\frac{\mathcal{A}'}{\mathcal{A}}\right|_{\rm fact}=1.16\pm0.18\:
\end{equation}
using LCSR form factors (see the discussion of Eq.~\eqref{A-fact}).
Consequently, the experimental determination of $|\mathcal{A}'/\mathcal{A}|$ is yet another
interesting topic for the LHCb upgrade. It will provide valuable insights into possible 
non-factorisable $U$-spin-breaking effects and the hadronisation dynamics of 
the \mbox{$B^0_{s,d}\to J/\psi K_{\rm S}^0$} system.

\boldmath
\section{$B$ Decays into Two Vector Mesons}\label{sec:psiV}
\unboldmath
\subsection{Preliminaries}
In the case of the $B^0_s\to J/\psi \phi$ and $B^0_d\to J/\psi \rho^0$ modes, the 
final states are mixtures of CP-even and CP-odd eigenstates. For the analysis of CP 
violation, these states have to be disentangled with the help of a time-dependent angular 
analysis of the $J/\psi\to\ell^+\ell^-$ and $\phi\to K^+K^-$, $\rho^0\to \pi^+\pi^-$ decay 
products \cite{DDLR,DDF}. To this end, it is convenient to introduce linear polarisation 
amplitudes $A_0(t)$, $A_\parallel(t)$ and $A_\perp(t)$ \cite{rosner}, where the 
$0$ and $\parallel$ final state configurations are CP-even while $\perp$ describes a 
CP-odd state. A detailed discussion of the general structure of the various observables
provided by the angular distribution in the presence of the penguin contributions was 
given in Ref.~\cite{RF-ang}. The linear polarisation states are also employed for the 
theoretical description of the $B^0_s\to J/\psi \Kstar$ decay, which has a flavour-specific
final state \cite{FFM}.
\boldmath
\subsection{The $B^0_s\to J/\psi \phi$ Channel}
\unboldmath
The decay \mbox{$B^0_s\to J/\psi \phi$} is the $B^0_s$-meson counterpart of 
\mbox{$B^0_d\to J/\psi K_{\rm S}^0$}. 
Assuming that the $\phi$ meson is a pure $s\bar{s}$ state, i.e.\ neglecting 
$\omega$--$\phi$ mixing (for a detailed discussion, see Ref.~\cite{FFM}),
this transition arises if we replace the down spectator quark of $B^0_d\to J/\psi K_{\rm S}^0$
by a strange quark. In analogy to Eq.~\eqref{Eq:DA_Bd2JpsiKS}, 
the SM decay amplitude takes the following form \cite{RF-ang,FFM}:
\begin{equation}\label{Bspsiphi-ampl}
A\left(B_s^0\rightarrow (J/\psi \phi)_f\right) = \left(1-\frac{\lambda^2}{2}\right)\mathcal{A}'_f
\left[1+\epsilon a'_fe^{i\theta'_f}e^{i\gamma}\right]\:,
\end{equation}
where the label $f\in\{0,\parallel,\perp\}$ distinguishes between the 
different configurations of the final state vector mesons.
We have to make the replacements
\begin{equation}
B^0_s\to J/\psi \phi: \,  b_f e^{i\rho_f} \, \rightarrow \, - \epsilon a'_f e^{i\theta'_f}\:,
\qquad
{\cal N}_f \, \rightarrow \,  \left(1-\frac{\lambda^2}{2}\right)\mathcal{A}'_f\:
\end{equation}
in order to apply the formalism introduced in Section~\ref{sec:hadr}.  
The hadronic phase shift
\begin{equation}
\phi_{s,(\psi \phi)_f}^{\rm eff} = \phi_s+\Delta\phi_s^{(\psi \phi)_f}
\end{equation}
can be obtained from Eq.~\eqref{Eq:DeltaPhi_Def}.

The penguin parameters $(a'_f,\theta'_f)$ are --- in general --- expected to differ for different 
final-state configurations $f$. However, applying simplified arguments along the lines of 
factorisation, the following picture emerges \cite{RF-ang}:
\begin{equation}\label{pen-rel}
a_f'\equiv a_{\psi\phi}'\:, \qquad \theta_f'\equiv \theta_{\psi\phi}'\:
\qquad  \forall f\in \{0,\parallel,\perp\}\:.
\end{equation}
The reason giving rise to the polarisation-independent parameters is the feature that form 
factors, which may depend on the final-state configuration $f$, cancel in the $a'_f$ ratios 
of penguin to tree amplitudes. It is an interesting question to test Eq.~\eqref{pen-rel} with 
experimental data, in particular in view of the discussion in the paragraph after 
Eqs.~(\ref{a-rel}) and (\ref{Aprime-A-rel}). The parameters $a_{\psi\phi}'$ and $\theta_{\psi\phi}'$ in 
Eq.~\eqref{pen-rel} may differ from their
$B^0_d\to J/\psi K_{\rm S}^0$ counterparts in Eq.~\eqref{Eq:DA_Bd2JpsiKS} due to the different 
hadronisation dynamics and non-factorisable effects.

The LHCb collaboration has recently presented the first results for the effective
\mbox{$B^0_s$--$\bar B^0_s$} mixing phases for the different final-state 
polarisations \cite{Aaij:2014zsa}:
\begin{alignat}{2}\label{phi_s0}
\phi_{s,0}^{\rm eff} & = -0.045\pm0.053\pm0.007 & \:= -(2.58 \pm  3.04  \pm0.40)^\circ\:, \\
\phi_{s,\parallel|}^{\rm eff}-\phi_{s,0}^{\rm eff} 
& = -0.018\pm0.043\pm0.009 & \:= -(1.03 \pm  2.46  \pm0.52)^\circ\:, \\
\phi_{s,\perp}^{\rm eff} -\phi_{s,0}^{\rm eff} 
& = -0.014\pm0.035\pm0.006 & \:= -(0.80 \pm  2.01  \pm0.34)^\circ\:.
\end{alignat}
Within the uncertainties, no dependence on the final-state configuration is revealed. 
Moreover, Eq.~\eqref{phi_s0} is in excellent agreement with the SM value in Eq.~\eqref{phis-SM}.
Using Eq.~\eqref{Dphi-fit} as a guideline for the size of possible hadronic 
phase shifts in $B^0_s\to J/\psi \phi$, the current precision is not yet high enough 
for resolving such effects. However, the LHCb analysis of Ref.~\cite{Aaij:2014zsa}
has a pioneering character, and it will be very interesting to monitor the polarisation-dependent
measurements as the precision increases. Assuming a universal value of $\phi_s^{\rm eff}$,
i.e.\ the relations in Eq.~\eqref{pen-rel},
the following result is obtained from the time-dependent analysis of the 
$B^0_s\to J/\psi[\to\mu^+\mu^-] \phi[\to K^+K^-]$ angular distribution \cite{Aaij:2014zsa}:
\begin{equation}\label{phis-eff-univ}
\phi_s^{\rm eff}=\phi_s + \Delta \phi_s =
-0.058\pm0.049\pm0.006 = -(3.32 \pm  2.81  \pm0.34)^\circ\:.
\end{equation}

The LHCb collaboration has also reported first polarisation-dependent results for the 
following quantities:
\begin{equation}\label{lam-expr}
|\lambda_f|\equiv \left|\frac{A(\bar B^0_s\to (J/\psi \phi)_f}{A(B^0_s\to (J/\psi \phi)_f}\right|=
\left|\frac{1+\epsilon a'_fe^{i\theta'_f}e^{-i\gamma}}{1+\epsilon a'_fe^{i\theta'_f}e^{+i\gamma}}\right|\:.
\end{equation}
In this expression, CP violation in $B^0_s$--$\bar B^0_s$ oscillations, which is a tiny 
effect \cite{Lenz}, has been neglected, like in Eq.~\eqref{ACP} with 
Eqs.~\eqref{AD-expr}, \eqref{AM-expr} and \eqref{Eq:ADG}. The LHCb measurements 
are given by
\begin{align}
|\lambda^0| & =1.012\pm0.058\pm0.013\:, \\
|\lambda^\perp/\lambda^0| & =1.02\pm0.12\pm0.05\:,\\
|\lambda^\parallel/\lambda^0| & =0.97\pm0.16\pm0.01\:.
\end{align}
Within the current uncertainties, again no polarisation dependence is observed. This is in 
agreement with the structure of Eq.~\eqref{lam-expr}, where the parameter $a'_fe^{i\theta'_f}$ 
enters with $\epsilon\sim0.05$. If we use the fit result in Eq.~\eqref{Eq:B2PV_chi2fit} as a guideline 
and assume $a'_fe^{i\theta'_f}\sim0.2$, we obtain $|\lambda^f|=1+{\cal O}(0.01)$, which sets
the scale of the required precision to resolve possible footprints of the penguin contributions 
in these measurements.

\begin{figure}[t]
\center
\includegraphics[width=0.75\textwidth]{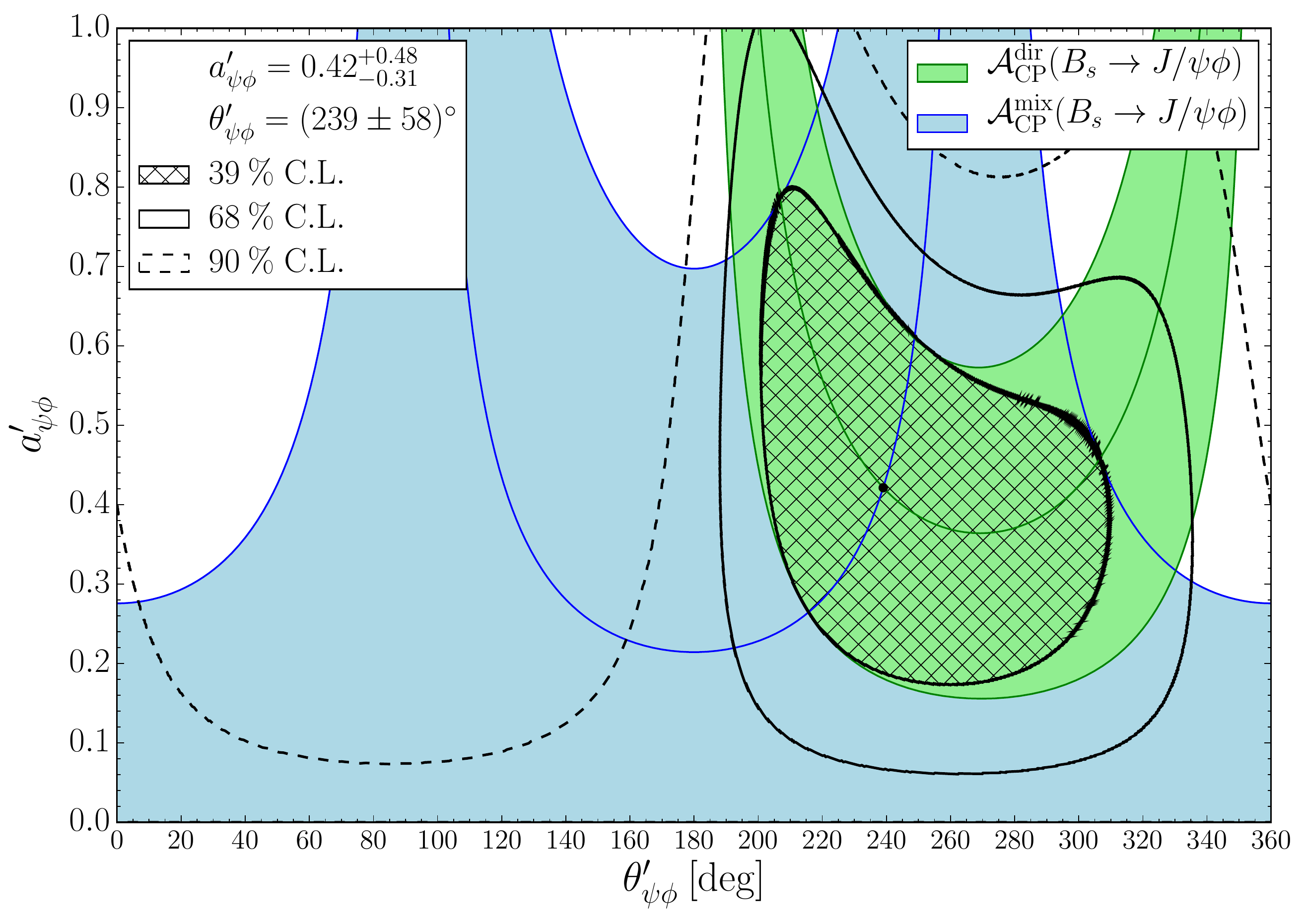}
\caption{Constraints in the $\theta'_{\psi\phi}$--$a'_{\psi\phi}$ plane following from the 
effective $B^0_s$--$\bar B^0_s$ mixing phase in Eq.~\eqref{phis-eff-univ} and 
$|\lambda_{\psi\phi}|$ in Eq.~\eqref{lam-exp-univ}. Here we assume the SM value of $\phi_s$ in 
Eq.~\eqref{phis-SM}.}\label{Fig:BsPsiPhi}
\end{figure}

Assuming that the parameters $|\lambda^f|\equiv |\lambda_{\psi\phi}|$ do not depend on the 
final-state configuration of the vector mesons, the LHCb collaboration has extracted the 
following result from the $B^0_s\to J/\psi[\to\mu^+\mu^-] \phi[\to K^+K^-]$ data:
\begin{equation}\label{lam-exp-univ}
|\lambda_{\psi\phi}|= 0.964\pm0.019\pm0.007\:.
\end{equation}
The deviation from unity at the $1.8\,\sigma$ level  --- which could well be an experimental
fluctuation --- would be surprisingly large in view of the discussion
given above. In Fig.~\ref{Fig:BsPsiPhi}, we convert this result into a contour band in the 
$\theta_{\psi\phi}'$--$a_{\psi\phi}'$ plane. As expected, the central value would correspond 
to penguin effects too large to be consistent with the other constraints. Assuming the SM value 
of $\phi_s$ in Eq.~\eqref{phis-SM}, we may also show the experimental result in 
Eq.~\eqref{phis-eff-univ} as a band in this figure. This analysis illustrates the observation 
we made in the context with Eqs.~\eqref{tan-phis} and \eqref{tan-phid}: in order to ensure 
a small phase shift of $\phi_s$ for large penguin parameters, strong phases  
around $\pm90^\circ$ are needed. Interestingly, the data for $B^0_d\to J/\psi \rho^0$ also 
suggest such a picture for the strong phases.

\begin{figure}[t]
\center
\includegraphics[width=0.45\textwidth]{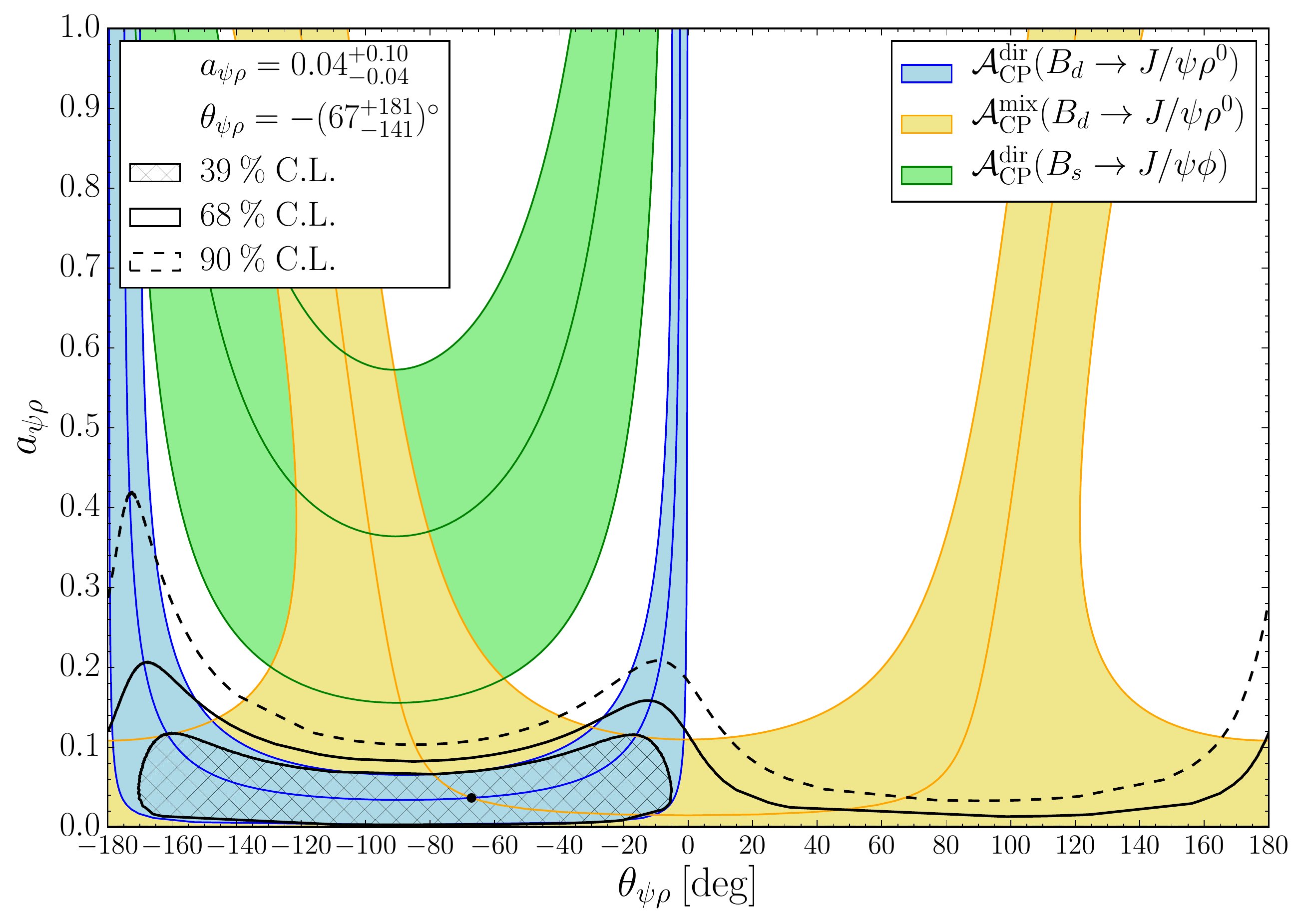}
\includegraphics[width=0.45\textwidth]{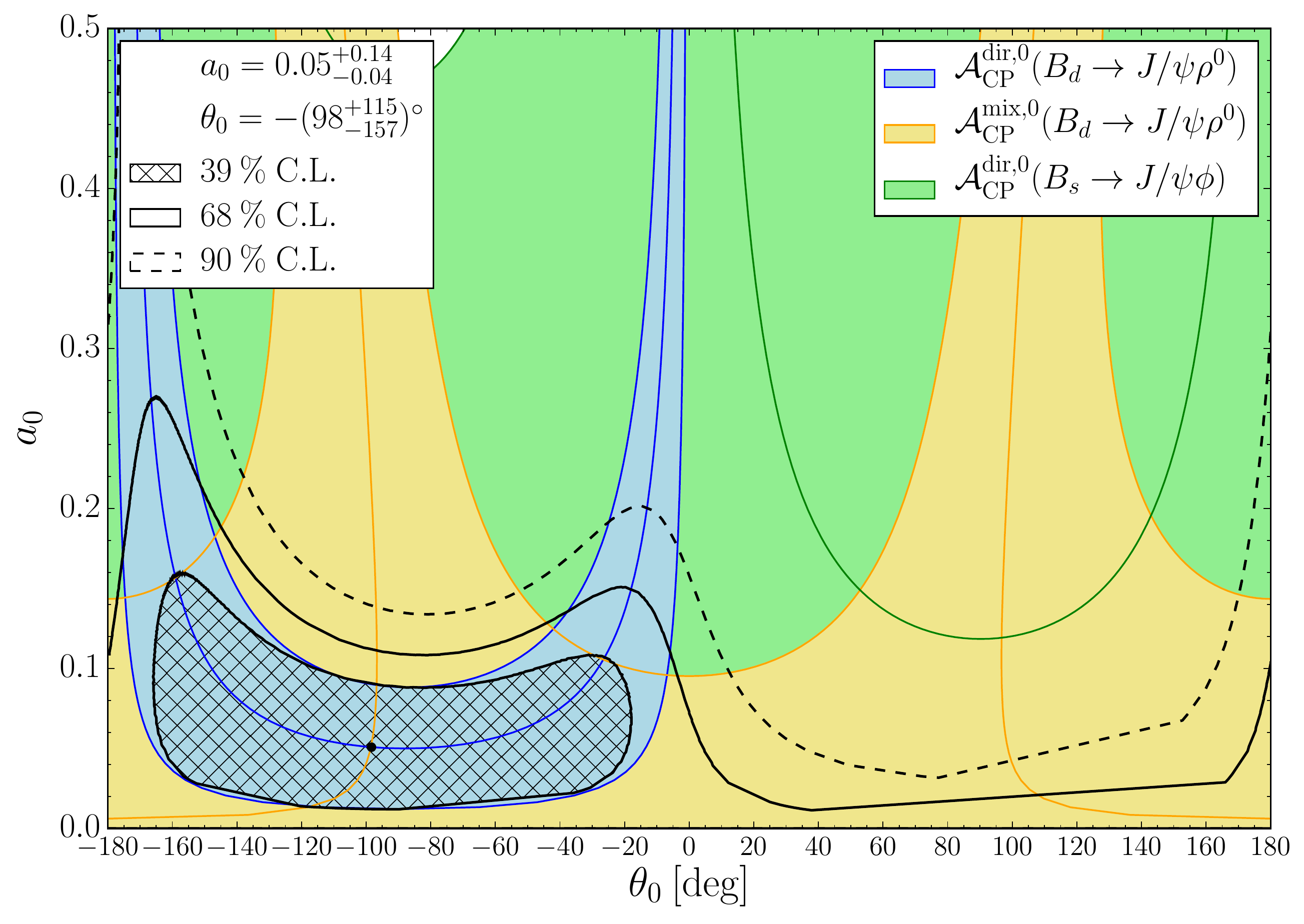}
\includegraphics[width=0.45\textwidth]{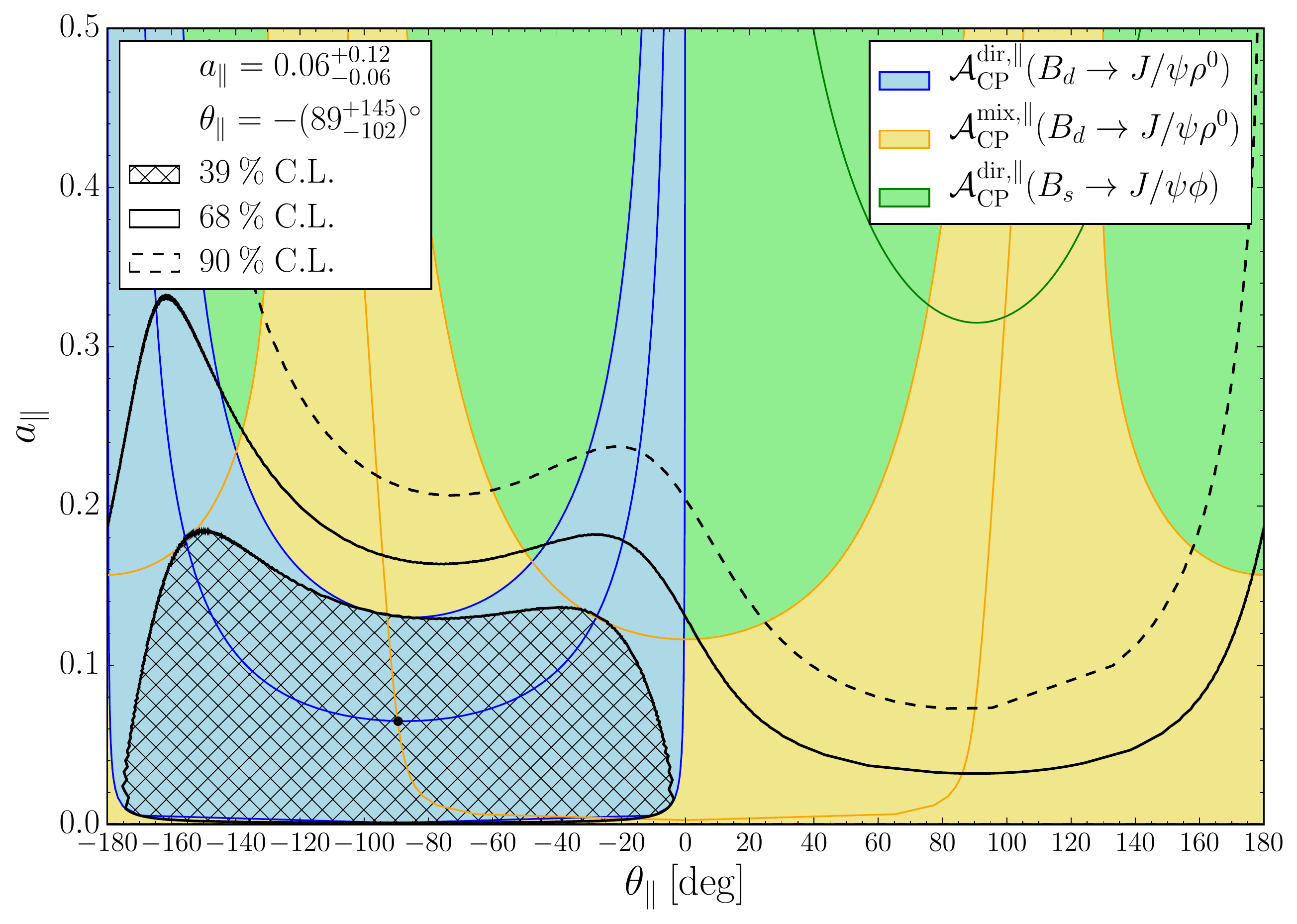}
\includegraphics[width=0.45\textwidth]{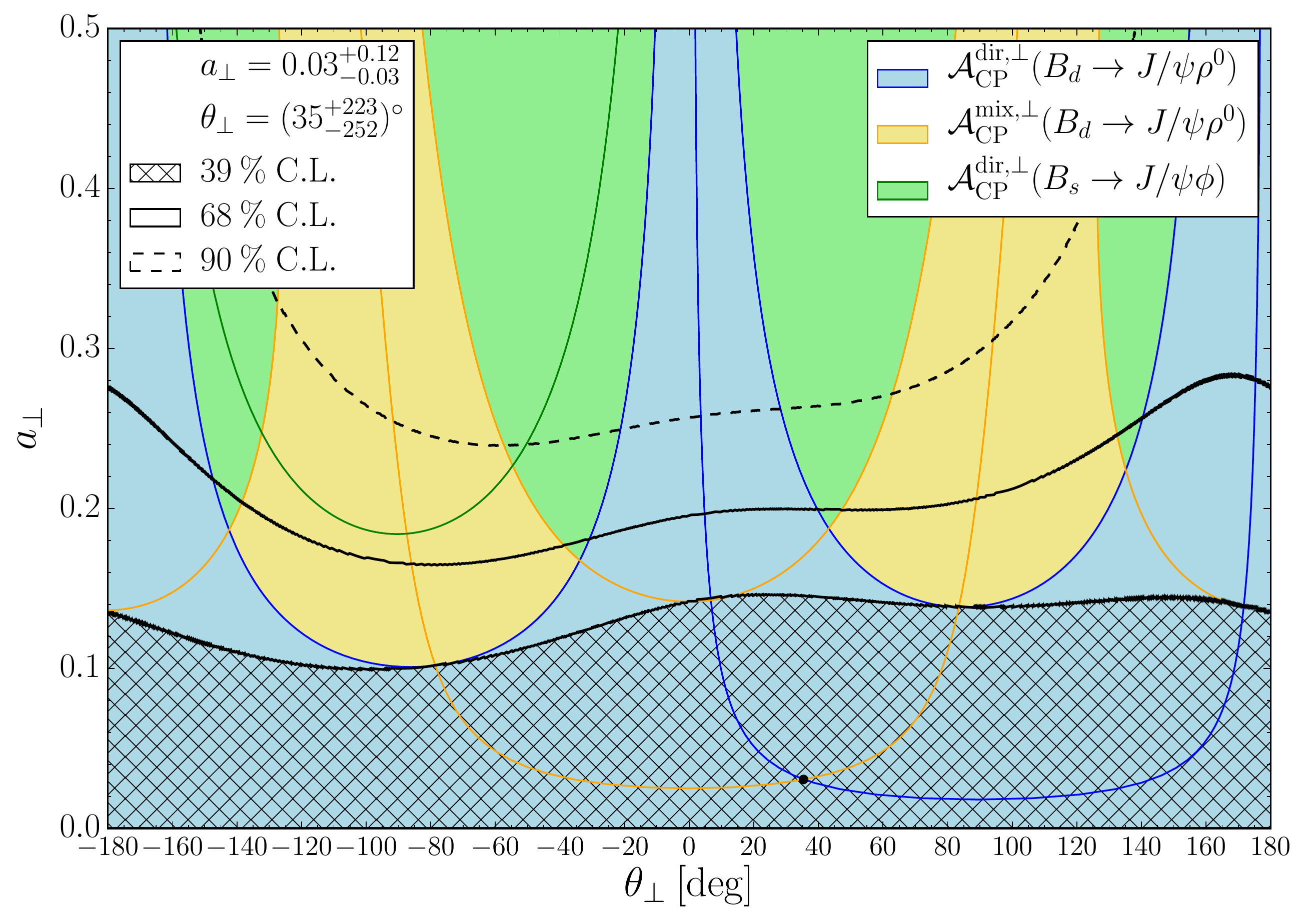}
\caption{Determination of the penguin parameters $a_f$ and $\theta_f$ from intersecting contours 
derived from the CP observables in \mbox{$B_d^0\to J/\psi\rho^0$.}
Superimposed are the confidence level contours obtained from a $\chi^2$ fit to the data.
The contour originating from the direct CP violation in \mbox{$B_s^0\to J/\psi\phi$} (see also
Fig.~\ref{Fig:BsPsiPhi}) has been added for visual comparison, but is not taken into account 
in the fit.}\label{Fig:B2VV}
\end{figure}

%
%
%
\boldmath
\subsection{The $B^0_d\to J/\psi \rho^0$ Channel}\label{Sec:B2JpsiRho}
\unboldmath
In analogy to $B^0_s\to J/\psi K_{\rm S}^0$ and $B^0_d\to J/\psi \pi^0$, the decay 
$B^0_d\to J/\psi \rho^0$ originates from $\bar b\to \bar d c \bar c$ quark-level transitions
and has a decay amplitude of similar structure \cite{RF-ang}:
\begin{equation}\label{Bpsirho-ampl}
\sqrt{2} \, A\left(B_d^0\rightarrow (J/\psi \rho^0)_f\right) = - \lambda \mathcal{A}_f
\left[1 - a_fe^{i\theta_f}e^{i\gamma}\right]\:,
\end{equation}
where the factor of $\sqrt{2}$ is due to the wave function of the $\rho^0$. In analogy to
$B^0_s\to J/\psi \phi$, the $B^0_d\to J/\psi \rho^0$ decay also shows mixing-induced 
CP violation, where an analysis of the $J/\psi \to \ell^+\ell^-$ and $\rho^0\to\pi^+\pi^-$
decay products is required to disentangle the CP-even and CP-odd final states. In order 
to apply the formalism of Section~\ref{sec:hadr}, we have to make the replacements
\begin{equation}
B^0_d\to J/\psi \rho^0: \,  b_f e^{i\rho_f} \, \rightarrow \, a_f e^{i\theta_f}, \qquad
{\cal N}_f \, \rightarrow \, - \frac{\lambda \mathcal{A}_f}{\sqrt{2}}\:.
\end{equation}
In particular, we then obtain expressions for the ``effective" mixing phases 
$\phi_{d,f}^{\rm eff}\equiv 2\beta^{\rm eff}_f$ by applying Eq.~\eqref{phiq-eff-def}. 
It should be emphasised that the corresponding penguin shifts are not doubly 
Cabibbo-suppressed. If we rescale the result in Eq.~\eqref{Dphi-fit}
by $-1/\epsilon$, we expect hadronic penguin shifts of ${\cal O}(20^\circ)$ in the 
$B^0_d\to J/\psi \rho^0$ channel. However, as we noted after Eqs.~\eqref{tan-phis} and 
\eqref{tan-phid}, and as we will see below, also the strong phases play an important role
for the numerical values. The hadronic parameters in the $B^0_{s,d}\to J/\psi K_{\rm S}^0$ 
system and in $B^0_d\to J/\psi \rho^0$ are generally expected to differ from one another. 

The LHCb collaboration has recently reported the first experimental results for CP 
violation in the $B^0_d\to J/\psi \rho^0$ channel \cite{LHCb-psi-rho}. The measurements of
the polarisation-dependent effective $B^0_d$--$\bar B^0_d$ mixing phases are given 
as follows:
\begin{align}
\phi_{d,0}^{\rm eff} & =\left(44.1\pm10.2^{+3.0}_{-6.9}\right)^\circ\:,\\
\phi_{d,\parallel}^{\rm eff}-\phi_{d,0}^{\rm eff} & =-\left(0.8\pm6.5^{+1.9}_{-1.3}\right)^\circ\:,\\
\phi_{d,\perp}^{\rm eff}-\phi_{d,0}^{\rm eff} & =-\left(3.6\pm7.2^{+2.0}_{-1.4}\right)^\circ\:.
\end{align}
Within the uncertainties, no dependence on the final-state configuration $f$ is detected. 
Assuming penguin parameters independent of $f$, i.e.\
\begin{equation}\label{pen-rel-rho}
a_f\equiv a_{\psi\rho}\:, \qquad \theta_f\equiv \theta_{\psi\rho}\:
\qquad  \forall f\in \{0,\parallel,\perp\}\:
\end{equation}
in analogy to the relations in Eq.~\eqref{pen-rel}, the phase 
\begin{equation}
\phi_{d}^{\rm eff}=\left(41.7\pm9.6^{+2.8}_{-6.3}\right)^\circ
\end{equation}
and the CP asymmetries
\begin{align}
{\cal A}_{\rm CP}^{\rm dir}(B_d\to J/\psi \rho)\equiv C_{J/\psi \rho} & 
=-0.063\pm0.056^{+0.019}_{-0.014}\:\\
- {\cal A}_{\rm CP}^{\rm mix}(B_d\to J/\psi \rho) \equiv S_{J/\psi \rho}& 
=-0.66^{+0.13+0.09}_{-0.12-0.03}\:
\end{align}
are extracted from the experimental analysis of the time-dependent angular distribution of the 
$B^0_d\to J/\psi [\to\mu^+\mu^-] \rho^0[\to \pi^+\pi^-]$ decay products.

The formulae in Section~\ref{sec:hadr} allow the conversion of these results into the 
$B^0_d\to J/\psi \rho^0$ penguin parameters. To this end,
we assume again the CKMfitter value of $\gamma$ in Eq.~\eqref{gamma-range}. 
Moreover, we need the $B^0_d$--$\bar B^0_d$ mixing phase $\phi_d$ as an input for the 
analysis of the mixing-induced CP asymmetry. 
However, as we have actually extracted $\phi_d$ from the global fit discussed in 
Section~\ref{ssec:constr}, we shall use the value in Eq.~\eqref{Eq:phid_chi2fit} for the 
$B^0_d\to J/\psi \rho^0$ analysis. In the LHCb study of Ref.~\cite{LHCb-psi-rho}, 
corrections from penguin contributions to $B^0_d\to J/\psi K_{\rm S}^0$ were not taken 
into account.

The main results of the $\chi^2$ fit to the data read as follows:
\begin{equation}\label{Eq:B2VV_chi2fit}
a_{\psi\rho} = 0.037^{+0.097}_{-0.037} \:,\qquad \theta_{\psi\rho} 
= -\left(67^{+181}_{-141}\right)^{\circ}\:,
\qquad \Delta\phi_d^{\psi\rho} = -\left(1.5_{-10}^{+12}\right)^{\circ}\:.
\end{equation}
In Fig.~\ref{Fig:B2VV}, we show the corresponding confidence level contours with the  
bands of the individual observables. It is interesting to note that the current experimental 
measurement of $|\lambda_{\psi\phi}|$ from \mbox{$B_s^0\to J/\psi\phi$} is in slight tension 
with the results from \mbox{$B^0_d\to J/\psi \rho^0$.} Should this turn out not to be a mere 
fluctuation of the data, which seems unlikely, the effect cannot be explained by penguin 
effects alone.

We have also explored a polarisation-dependent analysis of the penguin effects in 
\mbox{$B_s^0\to J/\psi\phi$} using the same strategy as for the above fit. 
The resulting confidence level contours are shown in Fig.~\ref{Fig:B2VV}. They 
are compatible with the polarisation-independent results in Eq.~\eqref{Eq:B2VV_chi2fit}, but
the current uncertainties are too large to draw further conclusions. This analysis should 
be seen as an illustration and motivation for experimentalists to perform more precise 
polarisation-dependent measurements, which are the method of choice in
the long run.

Neglecting exchange and penguin annihilation topologies (see Fig.~\ref{Fig:Feynman_Additional}), 
the $SU(3)$ flavour symmetry 
allows us to convert the hadronic parameters of the $B^0_d\to J/\psi \rho^0$ decay 
into their $B^0_s\to J/\psi \phi$ counterparts \cite{RF-ang}:
\begin{equation}\label{a-rel-VV}
a'_fe^{i\theta'_f}=a_f e^{i\theta_f}\:,\qquad \mathcal{A}'_f=\mathcal{A}_f\:,
\end{equation}
allowing us to convert the penguin parameters in Eq.~\eqref{Eq:B2VV_chi2fit} 
into the hadronic phase shift of the $B^0_s\to J/\psi \phi$ decay. Parametrising possible
$SU(3)$-breaking effects as in Eq.~\eqref{a-rel-break} with $\xi=1.00\pm0.20$ and 
$\delta=(0\pm 20)^\circ$, we obtain
\begin{equation}\label{phis-shift}
\Delta\phi_s^{\psi\phi} = \left[0.08_{-0.72}^{+0.56}\:(\text{stat})_{-0.13}^{+0.15}\:(SU(3))\right]^{\circ}\:,
\end{equation}
which is statistics limited, even when assuming larger $SU(3)$-breaking uncertainties.
The power of mixing-induced CP violation in $B^0_d\to J/\psi \rho^0$ for this determination
is remarkable \cite{LHCb-psi-rho}. It should be compared with the current value of 
$\phi_s^{\rm eff}$ in Eq.~\eqref{phis-eff-univ}, which is affected by significantly larger
experimental uncertainties. 

The contours in Fig.~\ref{Fig:B2VV} do not rely on information from decay 
rates and are theoretically clean. As in the discussion of the $B^0_s\to J/\psi K_{\rm S}^0$
benchmark scenario in Section~\ref{ssec:scene}, we may use 
the penguin parameters extracted from the CP asymmetries of $B^0_d\to J/\psi \rho^0$
to determine the ratio of CP-conserving strong amplitudes, in analogy to Eq.~\eqref{A-det}. 
The only conceptual difference is that polarisation-dependent studies should be performed
in the $B^0_d\to J/\psi \rho^0$ and $B^0_s\to J/\psi \phi$ systems. Following these lines,
we obtain the amplitude ratios
\begin{align}
\left|\frac{\mathcal{A}'_0(B_s\rightarrow J/\psi\phi)}{\mathcal{A}_0(B_d\rightarrow 
J/\psi \rho^0)}\right| & = 1.06 \pm 0.07\:(\text{stat}) \pm 0.04\:(a_0,\theta_0)\:\label{Arat-0}\\
\left|\frac{\mathcal{A}'_{||}(B_s\rightarrow J/\psi\phi)}{\mathcal{A}_{\parallel}(B_d\rightarrow 
J/\psi \rho^0)}\right| & = 1.08 \pm 0.08\:(\text{stat}) \pm 0.05\:(a_{\parallel},\theta_{\parallel})\:\\
\left|\frac{\mathcal{A}'_{\perp}(B_s\rightarrow J/\psi\phi)}{\mathcal{A}_{\perp}(B_d\rightarrow 
J/\psi \rho^0)}\right| & = 1.24 \pm 0.15\:(\text{stat}) 
\pm 0.06\:(a_{\perp},\theta_{\perp})\:,\label{Arat-perp}
\end{align}
which are still consistent with the limit of no  $SU(3)$-breaking corrections.
These results can be compared with QCD calculations, such as the recent results obtained
in Ref.~\cite{LWX} within the perturbative QCD (PQCD) approach. Within naive factorisation, 
the LCSR form factors of Ref.~\cite{Ball:2004rg} (see Table~8) yield
\begin{align}
\left|\frac{\mathcal{A}'_0(B_s\rightarrow J/\psi\phi)}{\mathcal{A}_0(B_d\rightarrow 
J/\psi \rho^0)}\right|_{\rm fact} & = 1.43 \pm 0.42\:\\
\left|\frac{\mathcal{A}'_{||}(B_s\rightarrow J/\psi\phi)}{\mathcal{A}_{\parallel}(B_d\rightarrow 
J/\psi \rho^0)}\right|_{\rm fact} & = 1.37 \pm 0.20\:\\
\left|\frac{\mathcal{A}'_{\perp}(B_s\rightarrow J/\psi\phi)}{\mathcal{A}_{\perp}(B_d\rightarrow 
J/\psi \rho^0)}\right|_{\rm fact} & = 1.25 \pm 0.15\:.
\end{align}
Although the uncertainties are still very large, these numbers are consistent with 
the results in Eqs.~\eqref{Arat-0}--\eqref{Arat-perp},
and imply
\begin{equation}
\left|\frac{\mathcal{A}'}{\mathcal{A}}\right| = \left|\frac{\mathcal{A}'_{\text{fact}}}{\mathcal{A}_{\text{fact}}}\right|
\left|\frac{1+\mathcal{A}'_{\text{non-fact}}/\mathcal{A}'_{\text{fact}}}{1+\mathcal{A}_{\text{non-fact}}/\mathcal{A}_{\text{fact}}}\right|
\approx  \left|\frac{\mathcal{A}'_{\text{fact}}}{\mathcal{A}_{\text{fact}}}\right|\:.
\end{equation}
Consequently, either the non-factorisable contributions $\mathcal{A}^{(\prime)}_{\text{non-fact}}$ themselves 
 or the difference (due to $SU(3)$-breaking effects) between the 
ratios $\mathcal{A}'_{\text{non-fact}}/\mathcal{A}'_{\text{fact}}$ and 
$\mathcal{A}_{\text{non-fact}}/\mathcal{A}_{\text{fact}}$ is small.
In view of the discussion after Eqs.~\eqref{a-rel} and \eqref{Aprime-A-rel}, the latter option is favoured.
A similar picture also arises for $SU(3)$-breaking effects in 
$B^0_d\to\pi^+\pi^-$, $B^0_d\to \pi^-K^+$, $B^0_s\to K^+K^-$ decays, which exhibit a
different decay dynamics \cite{RF-K-10}.
In view of this observation, we get confidence in
the first relation in Eq.~\eqref{a-rel-VV} (and the uncertainties assumed in 
Eq.~\eqref{a-rel-break}). It is interesting to note that the experimental 
uncertainties of the ratios in Eqs.~\eqref{Arat-0}--\eqref{Arat-perp} are already smaller 
or of similar size than the uncertainties of the theoretical calculations, which are challenging to improve.

\boldmath
\subsection{The $B^0_s\to J/\psi \Kstar$ Channel}
\unboldmath
The decay $B^0_s\to J/\psi \Kstar$ originates from $\bar b\to \bar d c \bar c$ quark-level 
processes and is the $B^0_s$-meson counterpart of the $B^0_d\to J/\psi \rho^0$ mode. 
The CDF \cite{Aaltonen:2011sy} and LHCb \cite{Aaij:2012nh} collaborations have 
measured the \mbox{$B^0_s\to J/\psi \Kstar$} branching ratio. In the SM, the 
decay amplitude takes the form
\begin{equation}\label{BpsiKast-ampl}
A(B_d^0\rightarrow (J/\psi \Kstar)_f) = - \lambda \mathcal{\tilde A}_f
\left[1 - \tilde a_fe^{i\tilde \theta_f}e^{i\gamma}\right]\:,
\end{equation}
where we have introduced the tilde to distinguish the hadronic $B^0_s\to J/\psi \Kstar$
parameters from their $B^0_d\to J/\psi \rho^0$ counterparts. Using $SU(3)$ flavour symmetry
arguments and neglecting penguin annihilation and exchange topologies in 
$B^0_d\to J/\psi \rho^0$, we obtain the relations
\begin{equation}\label{a-tilde-rel}
\tilde a_fe^{i\tilde \theta_f} = a_fe^{i\theta_f}\:,\qquad \mathcal{\tilde A}_f = \mathcal{A}_f\:.
\end{equation}
In order to apply the formalism of Section~\ref{sec:hadr}, we have to make the substitutions 
\begin{equation}
B^0_s\to J/\psi \Kstar: \,  b_f e^{i\rho_f} \, \rightarrow \, \tilde a_f e^{i\tilde \theta_f}\:, \qquad
{\cal N}_f \, \rightarrow \, - \lambda \mathcal{\tilde A}_f\:.
\end{equation}
In contrast to the $B^0_d\to J/\psi \rho^0$ channel, the $B^0_s\to J/\psi \Kstar$ decay 
does {\it not} exhibit mixing-induced CP violation as the $J/\psi \Kstar$  final state is 
flavour specific, i.e.\ the pion and kaon charges of $\Kstar \to \pi^+ K^-$ and 
$K^{*0} \to \pi^- K^+$ distinguish between initially present $B^0_s$ and $\bar B^0_s$ 
mesons, respectively. Consequently, in order to determine the penguin parameters, we 
have to rely on direct CP violation and decay rate information \cite{FFM}. For each of 
the final-state configurations $f\in\{0,\parallel,\perp\}$, we have a direct CP asymmetry 
$\mathcal{A}_{\rm CP}^{{\rm dir},f}$ and an observable corresponding to Eq.~\eqref{Eq:Hobs_Def}:
\begin{equation}
\tilde H_f\equiv  \frac{1}{\epsilon} \left|\frac{\mathcal{A}'_f}{\tilde{\mathcal{A}_f}}\right|^2
\frac{\text{PhSp}\left(B_s\rightarrow J/\psi \phi\right)}
{\text{PhSp}(B_s\rightarrow J/\psi \Kstar)}
\frac{\mathcal{B}(B_s\rightarrow J/\psi \Kstar)_{\text{theo}}}
{\mathcal{B}(B_s\rightarrow J/\psi \phi)_{\text{theo}}}
\frac{\tilde{f}_{\mathrm{VV},f}^{\rm exp}}{f_{\mathrm{VV},f}^{\rm exp}}\:,
\end{equation}
where
\begin{equation}
f_{\mathrm{VV},f}^{\rm exp} \equiv 
\frac{{\cal B}(B_s\to (f)_f)_{\rm exp}}{\sum_{f}{\cal B}(B_s\to (f)_f)_{\rm exp}}
\end{equation}
is the polarisation fraction of the $B_s\to f$ channel with $\sum_f f_{\mathrm{VV},f}^{\rm exp} =1$.
Also for the vector--vector modes the $H_f$ observables use the ``theoretical'' branching ratio 
concept, which for $B_s$ decays differs from the experimentally measured time-integrated 
branching ratio \cite{BR-paper}.
The conversion factors are similar to Eq.~\eqref{Eq:BR_correction} but become polarisation 
dependent. The measurement of these observables, which depend on $\tilde a_f$ 
and $\tilde \theta_f$ as well as $\gamma$, requires again an angular analysis of 
the decay products. Since $\gamma$ is an input, we may determine the 
penguin parameters for the different final state configurations $f$  \cite{FFM}.

In contrast to the analysis of $B^0_d\to J/\psi\rho^0$, where mixing-induced CP violation 
plays the key role, this method is affected by hadronic uncertainties which enter through 
the $\tilde H_f$ ratios. The extraction of these quantities from the data involves ratios of 
strong amplitudes, which depend on hadronic form factors and non-factorisable effects. 
In Ref.~\cite{LWX}, a detailed analysis of these quantities has been performed 
within the PQCD approach. We shall return to this topic below.

Measurements of direct CP asymmetries of the $B^0_s\to J/\psi \Kstar$ decay 
have not yet been performed. Using Eq.~\eqref{a-tilde-rel}, we expect them to equal 
those of $B^0_d\to J/\psi \rho^0$:
\begin{align}
\mathcal{A}_{\rm CP}^{\rm dir}(B_s\to J/\psi \Kstar)_{0} & = -0.094 \pm 0.071\:,\\
\mathcal{A}_{\rm CP}^{\rm dir}(B_s\to J/\psi \Kstar)_{\parallel} & = -0.12 \pm 0.12\:,\\
\mathcal{A}_{\rm CP}^{\rm dir}(B_s\to J/\psi \Kstar)_{\perp} & = \phantom{-}0.03 \pm 0.22\:,
\end{align}
where the CP asymmetries are defined as in Ref.~\cite{FFM}. It will be interesting to 
confront these numbers with future experimental results.

\section{Roadmap}\label{sec:road}
In the era of the LHCb upgrade and Belle II, there will be a powerful interplay of 
the different decay channels discussed in this paper. The measurement of CP violation 
in the $B^0_s\to J/\psi K_{\rm S}^0$ decay will allow us to extract the corresponding
penguin parameters in a theoretically clean way at LHCb and to control the penguin effects 
in the extraction of $\phi_d$ from $B^0_d\to J/\psi K_{\rm S}^0$ with the help of the $U$-spin 
symmetry \cite{RF-psiK,dBFK}. 

At Belle II, it will be important to measure CP violation in $B^0_d\to J/\psi \pi^0$ and to 
resolve the current discrepancy between the BaBar and Belle measurements of the 
mixing-induced CP asymmetry (see Eq.~\eqref{jpsipi-meas-mix}). The penguin parameters 
can be determined in analogy to the $B^0_s\to J/\psi K_{\rm S}^0$ strategy \cite{CPS}. However,
whereas the $U$-spin symmetry is sufficient in the case of the $B^0_{s,d}\to J/\psi K_{\rm S}^0$
system, the $B^0_d\to J/\psi \pi^0$ mode is affected by further uncertainties due to penguin 
annihilation and exchange topologies, which arise in $B^0_d\to J/\psi \pi^0$ but have 
no counterpart in $B^0_d\to J/\psi K_{\rm S}^0$. Some of these amplitudes are isospin suppressed 
(and thus expected to be very small) but
not those competing with the penguin contributions. The annihilation and exchange topologies
can be probed through $B^0_s\to J/\psi \pi^0$ \cite{FFJM}.
The LHCb collaboration does not see any evidence for the $B^0_s\to J/\psi \rho^0$ channel 
in the current data  \cite{Aaij:2014emv}.

Following these lines, the $B^0_d$--$\bar B^0_d$ mixing phase $\phi_d$ can be extracted
with unprecedented precision. The key question is whether the comparison with the
SM value $\phi_d^{\rm SM}=2\beta$ will result in a discrepancy, thereby indicating a 
CP-violating NP phase $\phi_d^{\rm NP}$. Here the interplay between $\gamma$ and the 
side $R_b$ of the UT is crucial \cite{BF-rev}:
\begin{equation}\label{s2b-expr}
\sin2\beta=\frac{2 R_b \sin\gamma(1-R_b\cos\gamma)}{(R_b \sin\gamma)^2+
(1-R_b\cos\gamma)^2}\:.
\end{equation}
The precision will be governed by $R_b$ \cite{Charles:2011va,Bevan:2014cya}.
Future data collected at the Belle II experiment and theoretical progress will hopefully 
resolve the discrepancy between the determination of $R_b$ from inclusive and exclusive 
semileptonic $B$ decays \cite{Agashe:2014kda}. The angle $\gamma$ can be determined
with high precision from $B\to D^{(*)}K^{(*)}$ decays, as given in Eq.~\eqref{gamma-LHCbupgrade}.
\newline

Measurements of the CP violation in $B^0_s\to J/\psi \phi$ will play a
key role for the determination of $\phi_s$. It will be important to have polarisation-dependent 
analyses of $\phi_{s,f}^{\rm eff}$ available with a precision much higher than 
the pioneering LHCb results reported recently in Ref.~\cite{Aaij:2014zsa}. Different values 
would signal the presence of penguin effects and a violation of the relations for the 
penguin parameters in Eq.~\eqref{pen-rel}. Measurements of the direct and mixing-induced 
CP-violating observables of the $B^0_d\to J/\psi \rho^0$ channel allow us to determine 
the corresponding penguin parameters in a clean way \cite{RF-ang}. Here the value of 
$\phi_d$ determined from the $B^0_{d,s}\to J/\psi K_{\rm S}^0$ system  
is needed as an input. Also in the $B^0_d\to J/\psi \rho^0$ analysis it will be important to 
make final-state-dependent measurements. The experimental results of 
Ref.~\cite{LHCb-psi-rho} provide a fertile ground for these analyses. Using then the relations
in Eq.~\eqref{a-rel-VV} allows us to determine the phase shifts $\Delta\phi_s^f$ and to 
extract the values of $\phi_s$ from the effective mixing phases $\phi_{s,f}^{\rm eff}$ of the
$B^0_s\to J/\psi \phi$ channel.

\begin{figure}[t]
\center
\includegraphics[width=1.02\textwidth]{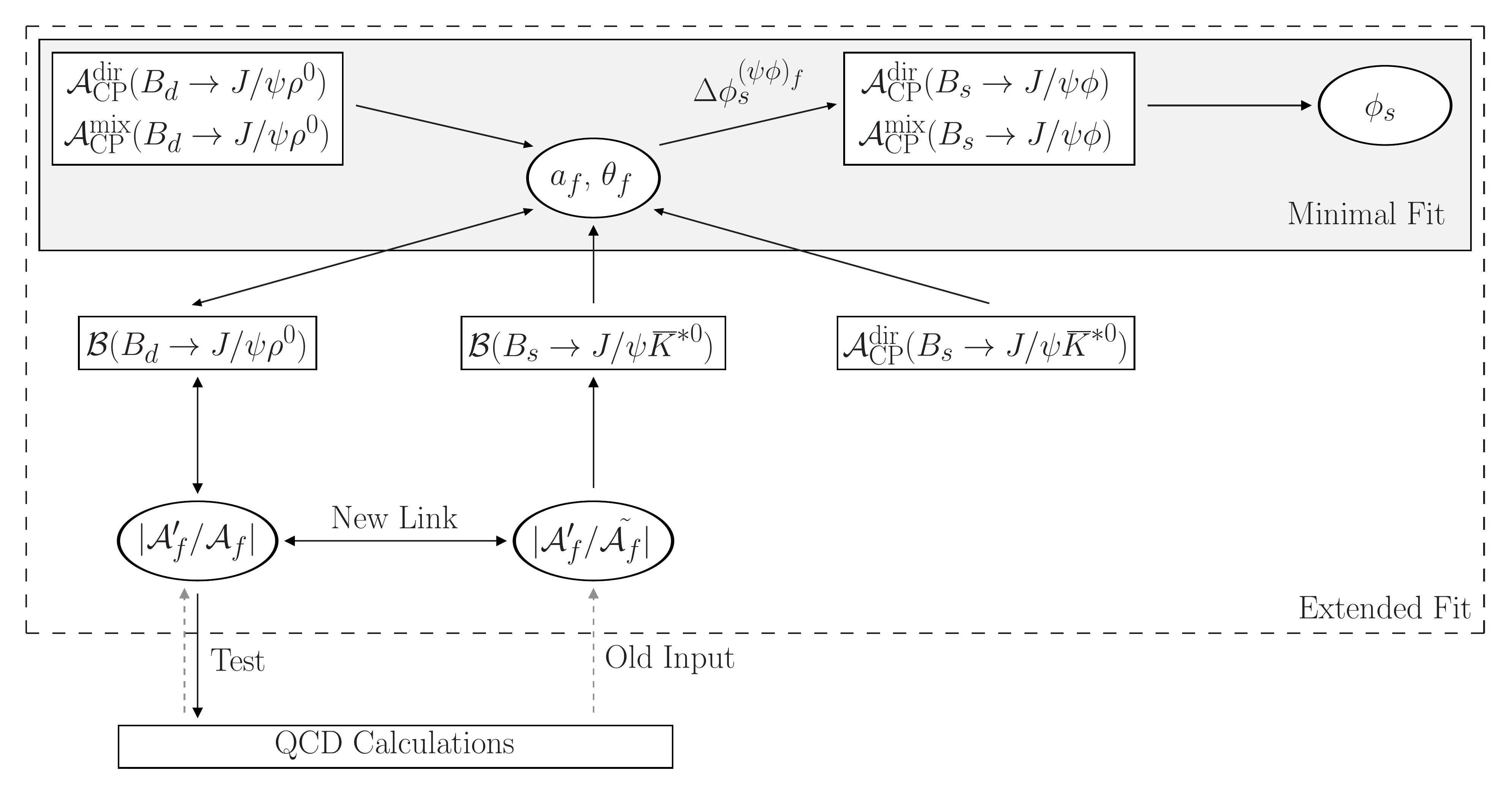}
\caption{Flow chart of the combined analysis of the $B_d^0\to J/\psi \rho^0$,  
$B^0_s\to J/\psi \Kstar$ and $B^0_s\to J/\psi \phi$ modes to simultaneously
determine the penguin parameters, the ratio of $SU(3)$-breaking strong amplitudes,
and the CP-violating $B^0_s$--$\bar B^0_s$ mixing phase $\phi_s$.}
\label{Fig:big-fit-flow}
\end{figure}

The penguin effects can also be probed by the $B^0_s\to J/\psi \Kstar$ decay \cite{FFM}.
This channel provides direct CP asymmetries but no mixing-induced CP violation as the final 
state is flavour-specific. In order to make use of the branching ratio information, ratios of 
strong amplitudes $|\mathcal{A}_f'/\mathcal{\tilde A}_f|$ are needed which introduce hadronic 
form-factor and non-factorisable uncertainties into the analysis. However, these ratios can actually 
be fixed through experiment. From the $B^0_d\to J/\psi \rho^0$, 
$B^0_s\to J/\psi \phi$ analysis, we may determine the ratios $|\mathcal{A}_f'/\mathcal{A}_f|$ 
in a theoretically clean way, as we discussed in Eqs.~\eqref{Arat-0}--\eqref{Arat-perp}
for the current data. Using the relation in Eq.~\eqref{a-tilde-rel}, we obtain
\begin{equation}
\left|\frac{\mathcal{A}_f'}{\mathcal{A}_f}  \right| = \left|\frac{\mathcal{A}_f'}{\mathcal{\tilde A}_f}  \right|,
\end{equation}
which allows us to convert the $B^0_s\to J/\psi \Kstar$ rate measurements into the
$\tilde H_f$ observables.  Finally, using also the relation 
\begin{equation}
\tilde a_fe^{i\tilde \theta_f} = a_fe^{i\theta_f}=a_f'e^{i\theta_f'},
\end{equation}
it is possible to make a simultaneous $\chi^2$ fit to the experimental data offered by the 
$B^0_s\to J/\psi \phi$, $B^0_d\to J/\psi \rho^0$, $B^0_s\to J/\psi \Kstar$ system
as illustrated in the flow chart in Fig.~\ref{Fig:big-fit-flow}. This global analysis allows 
us to combine all the information offered by the penguin control channels in an optimal 
way and provides valuable insights into strong interactions as a by-product.
Even though the direct CP asymmetry measurements in $B^0_s\to J/\psi \Kstar$ are at present not yet available,
we can already implement this strategy and extend the fits in Fig.~\ref{Fig:B2VV} to include branching ratio
information from $B^0_s\to J/\psi \phi$, $B^0_d\to J/\psi \rho^0$ and $B^0_s\to J/\psi \Kstar$.
The results of this analysis are
\begin{align}
\left|\frac{\mathcal{A}'_0}{\mathcal{A}_0}\right| & = 1.073_{-0.073}^{+0.094}\:,
& a_0 & = 0.05_{-0.04}^{+0.14}\:, & \theta_0 & = -\left(98_{-157}^{+115}\right)^{\circ}\:,\\
\left|\frac{\mathcal{A}'_{||}}{\mathcal{A}_{\parallel}}\right| & = 1.088_{-0.085}^{+0.114}\:,
& a_{||} & = 0.06_{-0.06}^{+0.12}\:, & \theta_{||} & = -\left(89_{-102}^{+145}\right)^{\circ}\:,\\
\left|\frac{\mathcal{A}'_{\perp}}{\mathcal{A}_{\perp}}\right| & = 1.21_{-0.13}^{+0.18}\:,
& a_{\perp} & = 0.03_{-0.03}^{+0.12}\:, & \theta_{\perp} & = \phantom{-}\left(35_{-252}^{+223}\right)^{\circ}\:.
\end{align}
We observe that with the current experimental precision the additional branching ratio information does not have
any impact on the determination of $a_f$ and $\theta_f$ with respect to the fits to the $B^0_d\to J/\psi \rho^0$
system only.
The information is fully used to constrain the amplitude ratios $|\mathcal{A}'_f/\mathcal{A}_f|$,
which were previously not included in the fit.
To observe any impact on $a_f$ and $\theta_f$, the combined experimental precision on the $H$ observables
needs to be improved by at least an order of magnitude.
Numerical differences in $|\mathcal{A}'_f/\mathcal{A}_f|$ compared to Eqs.~\eqref{Arat-0}--\eqref{Arat-perp} arise
due to the added information originating from the $B^0_s\to J/\psi \Kstar$ system.
This extended fit may be further refined by adding information from $B^0_s\to J/\psi \rho^0$ to probe 
exchange and penguin annihilation topologies.

\begin{figure}[t]
\center
\includegraphics[width=0.4\textwidth]{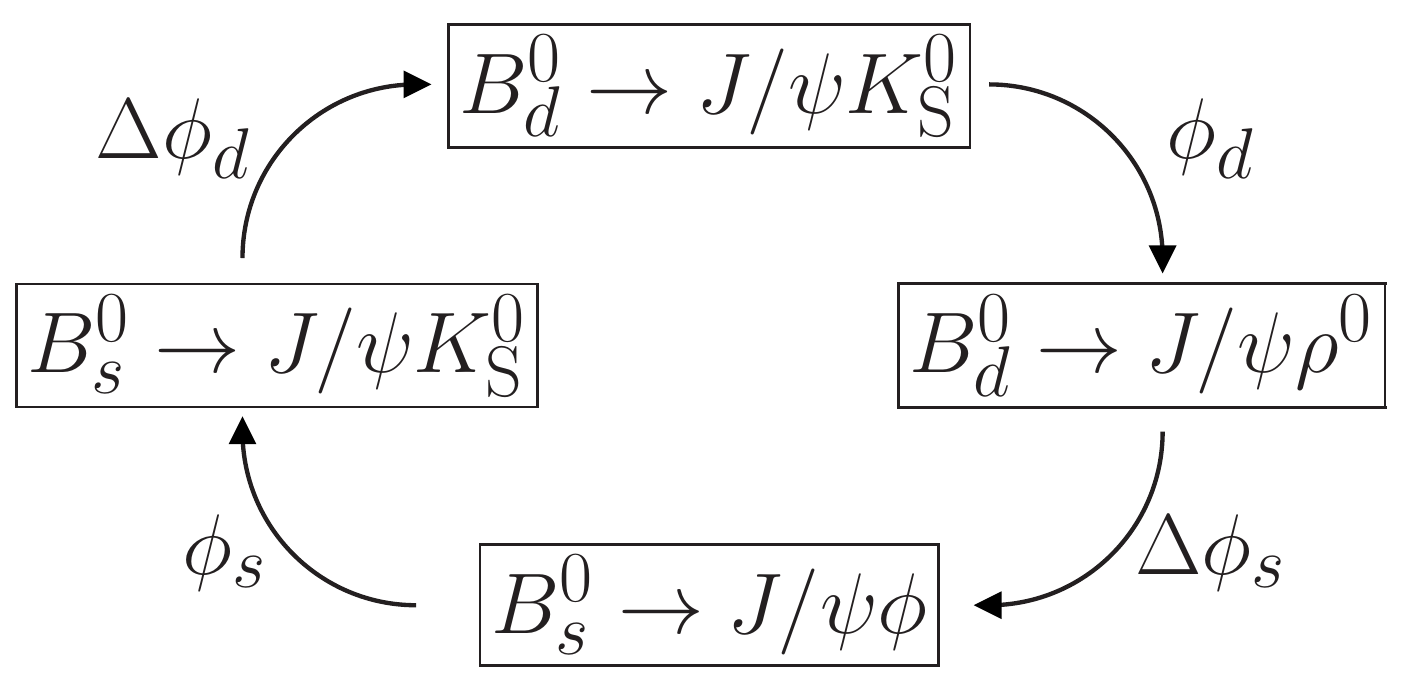}
\caption{Interplay between the decays used to measure the $B_q^0$--$\bar{B}_q^0$ mixing 
phases and the channels needed to control the penguin contributions in the former measurements.}
\label{Fig:Interplay}
\end{figure}

There is actually an interplay between the high-precision determinations of $\phi_d$ and $\phi_s$. 
The point is that $\phi_d$ is needed as an input for the analysis of mixing-induced CP 
violation of $B^0_d\to J/\psi \rho^0$ whereas $\phi_s$ is required for the analysis of
mixing-induced CP violation of $B^0_s\to J/\psi K_{\rm S}^0$. We have illustrated these
cross links in Fig.~\ref{Fig:Interplay}. Consequently, it will be advantageous to eventually perform
a simultaneous analysis of the $B^0_{s,d}\to J/\psi K_{\rm S}^0$ and 
$B^0_s\to J/\psi \phi$, $B^0_d\to J/\psi \rho^0$, $B^0_s\to J/\psi \Kstar$ systems.

\begin{figure}[t]
\center
\includegraphics[width=0.45\textwidth]{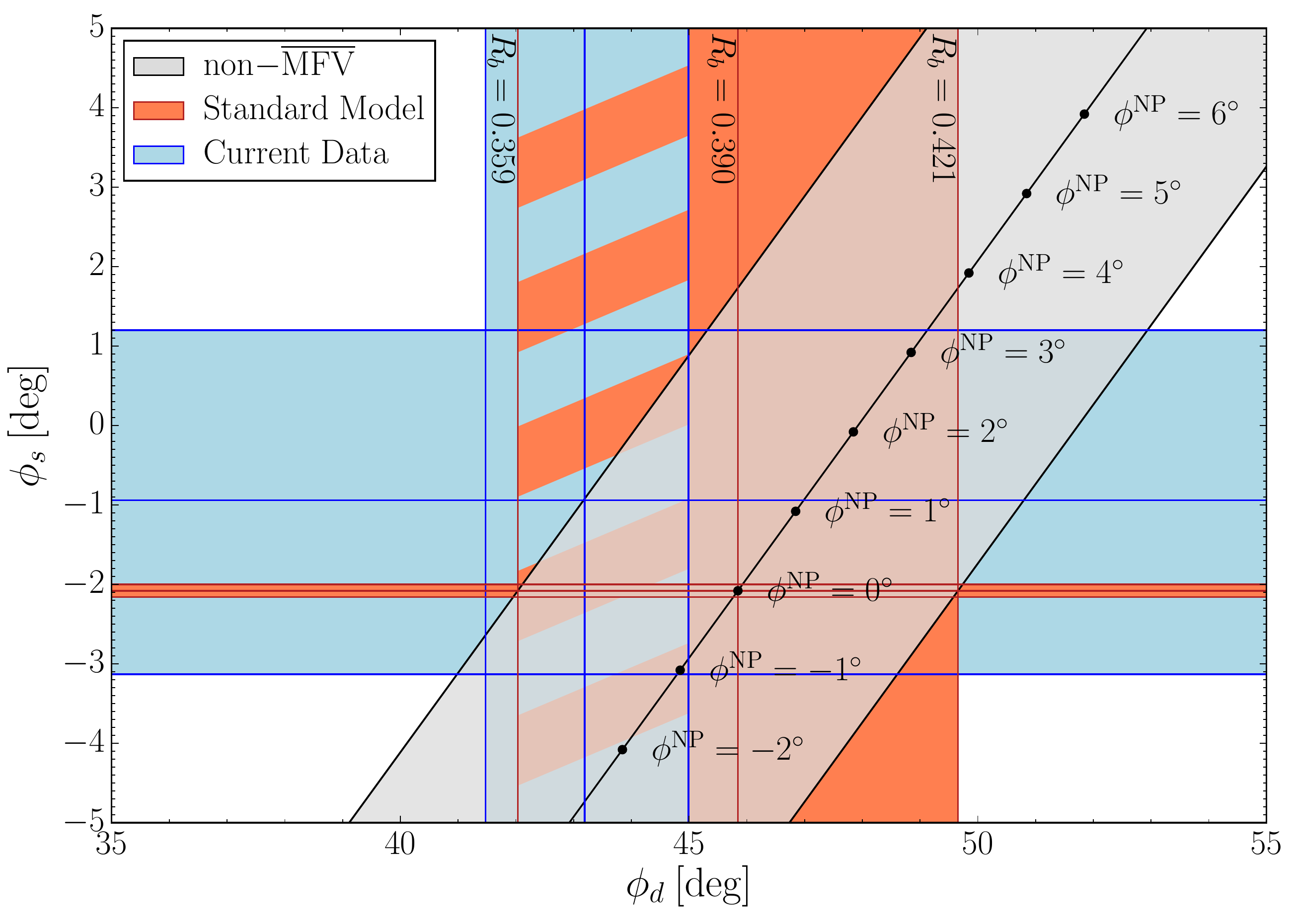}
\hfill
\includegraphics[width=0.45\textwidth]{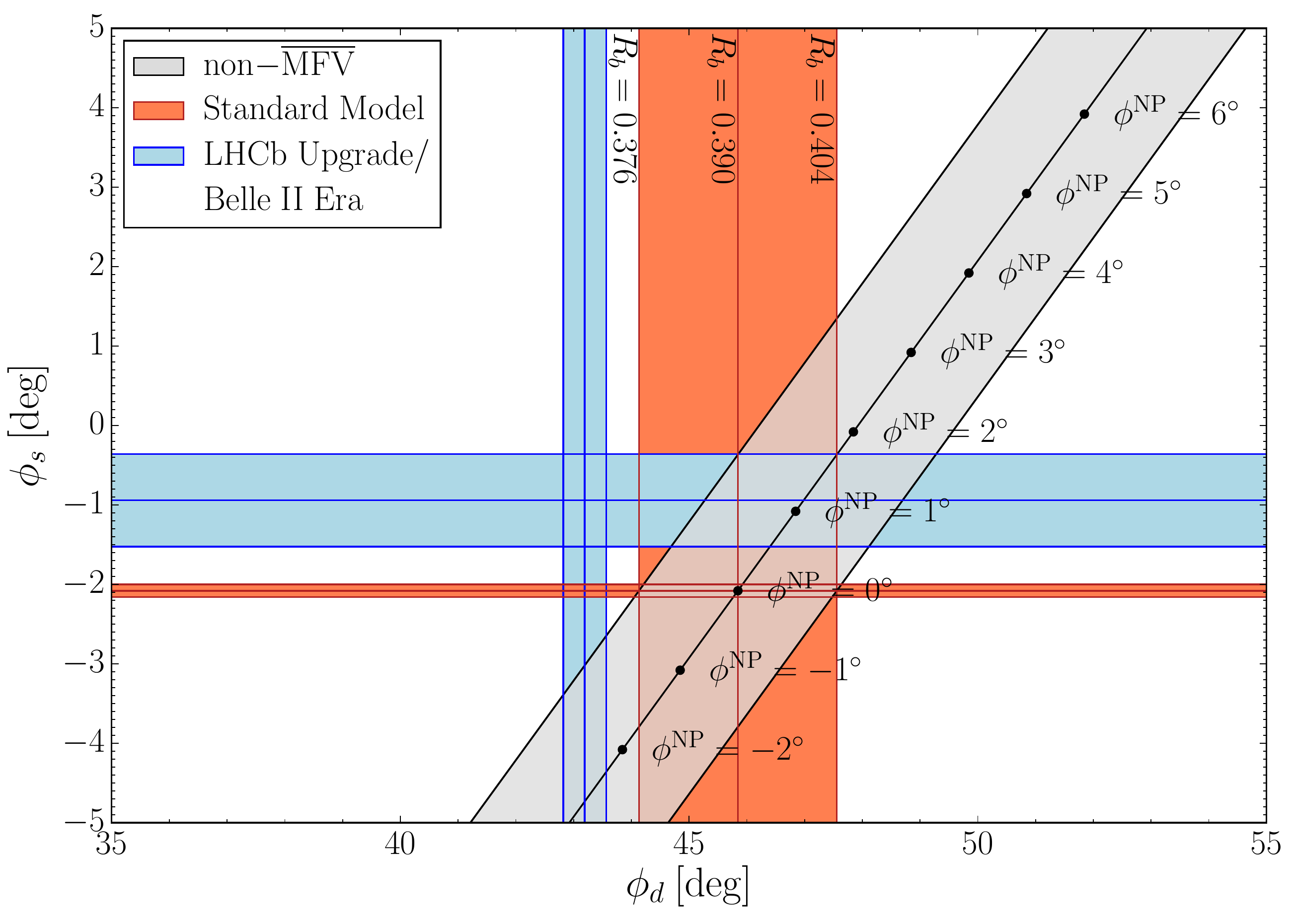}
\caption{Illustration of the correlation between $\phi_s$ and $\phi_d$ for 
$\mbox{non-}\overline{\mbox{MFV}}$ models with flavour-universal CP-violating 
NP phases characterised by Eq.~\eqref{non-MFV-rel}: we show the current 
experimental situation (left) and extrapolate to the LHCb upgrade era (right).}\label{Fig:non-MFV}
\end{figure}

For the search of NP in the era of the LHCb upgrade \cite{LHCb-implications}
and Belle II \cite{Abe:2010gxa}, it will be 
important to have determinations of both $\phi_d$ and $\phi_s$ available with the highest possible
precision. We obtain an interesting correlation between these mixing phases 
if their NP phases in Eq.~\eqref{mix-phases} take the same value:
\begin{equation}
\phi_s^{\rm NP}=\phi_d^{\rm NP}\equiv \phi^{\rm NP}.
\end{equation}
This relation, which was considered in Refs.~\cite{BaFl,BuGa} on a phenomenological 
basis, arises actually in extensions of the SM going beyond ``minimal flavour violation" 
(MFV), which are characterised by flavour-universal CP-violating NP
phases (for an overview, see Ref.~\cite{BurGir}). In this specific class of NP, 
referred to as  $\mbox{non-}\overline{\mbox{MFV}}$ models, we obtain the 
following correlation:
\begin{equation}\label{non-MFV-rel}
\phi_s = \phi_d +\left(\phi_s^{\rm SM}-\phi_d^{\rm SM}\right),
\end{equation}
which allows an experimental test. In Fig.~\ref{Fig:non-MFV}, we illustrate this relation both
for the current situation and for the expected situation in the LHCb upgrade era. 
The future uncertainty of the value of $\phi_d^{\rm SM}=2\beta$ will
be fully governed by $R_b$ (see Eq.~(\ref{s2b-expr})), which enters also the band 
representing the relation in Eq.~\eqref{non-MFV-rel}. It will be interesting to confront 
these considerations with experimental data in the next decade.

\section{Conclusions}\label{sec:concl}
The picture emerging from run I of the LHC suggests that we have to prepare ourselves to 
deal with smallish NP effects. For the determination of the $B_q^0$--$\bar{B}_q^0$ mixing 
phases $\phi_d$ and $\phi_s$ from CP violation measurements in $B_d^0\to J/\psi K_{\rm S}^0$ 
and $B_s^0\to J/\psi\phi$, respectively, this implies that controlling higher order hadronic 
corrections, originating from doubly Cabibbo-suppressed penguin topologies, becomes 
mandatory. In this paper, we have outlined strategies to accomplish this task using the 
$SU(3)$ flavour symmetry of QCD.

The penguin contributions to $B_d^0\to J/\psi K_{\rm S}^0$ can be controlled with the help 
of its $U$-spin partner $B_s^0\to J/\psi K_{\rm S}^0$. As the required CP violation measurements 
of the latter mode are not yet available, we have performed a global fit to current data 
for CP asymmetries and branching ratios of $B\to J/\psi (\pi/K)$ modes with similar 
dynamics to already constrain the hadronic penguin shift affecting the 
$B_d^0\to J/\psi K_{\rm S}^0$ channel. For the future 
LHCb upgrade era we have illustrated the potential of the $B_s^0\to J/\psi K_{\rm S}^0$ mode, 
which represents the cleanest penguin probe, with a benchmark scenario. In addition, we have
discussed a strategy to probe non-factorisable $U$-spin-breaking effects in the
$B_{s,d}^0\to J/\psi K_{\rm S}^0$ system.

The penguin contributions to $B_s^0\to J/\psi\phi$ can be controlled with the help of the 
modes $B_d^0\to J/\psi \rho^0$ and $B_s^0\to J/\psi \Kstar$. We have analysed the first 
LHCb measurement of CP violation in $B_d^0\to J/\psi \rho^0$, taking into account possible 
penguin effects in the required input for $\phi_d$. In view of the excellent precision that can 
already be obtained in this analysis, the $B_d^0\to J/\psi \rho^0$ mode is expected to play the
key role for the control of the penguin effects in the determination of $\phi_s$. We have
proposed a new strategy to add the $B_s^0\to J/\psi \Kstar$ data to this analysis in a 
global fit, which does not require knowledge of form factors for the
interpretation of the decay rate information. It rather allows us to determine also hadronic
parameters, which then provide insights into non-factorisable $SU(3)$-breaking effects. Adding
$B^0_s \to J/\psi \rho^0$ to the analysis, also the impact of penguin annihilation and exchange
topologies, which are expected to be small, can be probed through experimental data. 

Finally, we propose a combined analysis of the $B_{s,d}^0\to J/\psi K_{\rm S}^0$
and $B_s^0\to J/\psi\phi$, $B_d^0\to J/\psi \rho^0$, 
$B_s^0\to J/\psi \Kstar$ systems in order to simultaneously determine the mixing phases 
$\phi_d$ and $\phi_s$, taking into account the cross-correlations between these modes in
the control of the penguin effects. For the search of new sources of CP violation in the era of the
LHCb upgrade and Belle II, simultaneous high-precision measurements of $\phi_d$ and $\phi_s$ 
are crucial ingredients. In extensions of the SM, such as $\mbox{non-}\overline{\mbox{MFV}}$ 
models, characteristic correlations between $\phi_d$ and $\phi_s$ arise which can then be 
tested. While the SM prediction of $\phi_s$ has already a precision 
much smaller than the LHCb upgrade sensitivity, the major limitation for $\phi_d^{\rm SM}$  is 
given by the determination of $|V_{ub}/V_{cb}|$ entering the UT side 
$R_b$. Future progress on this long-standing challenge would be very desirable to complement
the cutting-edge analyses of CP violation. We look forward to moving to the high-precision frontier!

\subsubsection*{Acknowledgements}
We would like to thank Patrick Koppenburg for discussions and comments on the 
manuscript and are grateful to Sheldon Stone for correspondence.

\appendix
\section{Contributions from Annihilation Topologies}\label{App:annihil}
The framework introduced in Section \ref{sec:hadr} can be extended to allow for annihilation 
topologies $A_c$. The amplitude of the decay $B^+\to J/\psi\pi^+$
can be written as
\begin{equation}
A(B^+\to J/\psi\pi^+)=  - \lambda \mathcal{A}_{\rm c}\left[1- a_{\rm c} e^{i\theta_{\rm c}}
e^{i\gamma}\right],
\end{equation}
where
\begin{equation}
\mathcal{A}_{\rm c} \equiv \lambda^2 A \left[C_{\rm c}+P_{\rm c}^{(c)}-P_{\rm c}^{(t)}\right]
\end{equation}
is defined as in Eq.~(\ref{Aprime-def}), whereas
\begin{equation}
a_{\rm c} e^{i\theta_{\rm c}}=\tilde a_{\rm c} e^{i\tilde \theta_{\rm c}} + x e^{i\sigma}
\end{equation}
with
\begin{equation}
\tilde a_{\rm c} e^{i\tilde \theta_{\rm c}}\equiv R_b\left[\frac{P_{\rm c}^{(u)}-P_{\rm c}^{(t)}}{C_{\rm c}
+P_{\rm c}^{(c)}-P_{\rm c}^{(t)}}\right]
\end{equation}
and 
\begin{equation}
x e^{i\sigma} \equiv R_b\left[\frac{A_c}{C_{\rm c} + P_{\rm c}^{(c)} - P_{\rm c}^{(t)}}\right].
\end{equation}
The penguin parameter $\tilde a_{\rm c} e^{i\tilde \theta_{\rm c}}$ is defined in analogy to
Eq.~\eqref{Eq:Penguin_Def}, while the relative contribution from the annihilation 
topology is probed by $x e^{i\sigma}$. 
The direct CP asymmetry in \mbox{$B^+\to J/\psi\pi^+$} then takes the form
\begin{equation}
\mathcal{A}_{\text{CP}}^{\text{dir}} = \frac{2(\tilde a_{\rm c}\sin\tilde \theta_{\rm c} + 
x\sin\sigma)\sin\gamma}
{1 - 2(\tilde a_{\rm c}\cos\tilde \theta_{\rm c}+  x\cos\sigma)\cos\gamma 
+ 2\tilde a_{\rm c} x\cos(\tilde \theta_{\rm c}-\sigma)+ \tilde a_{\rm c}^2+x^2 }\:,
\end{equation}
whereas the ratio \mbox{$\Xi\left(B^{\pm}\to J/\psi \pi^{\pm},B_s\to J/\psi K_{\mathrm S}^0\right)$} depends on $x$ and $\sigma$ as
\begin{equation}
\Xi = \frac{1 - 2(\tilde a_{\rm c}\cos\tilde \theta_{\rm c}+  x\cos\sigma)\cos\gamma 
+ 2\tilde a_{\rm c} x\cos(\tilde \theta_{\rm c}-\sigma)+ \tilde a_{\rm c}^2+x^2}
{1 - 2\tilde a_{\rm c}\cos\tilde\theta_{\rm c}\cos\gamma + \tilde a_{\rm c}^2}\:.
\end{equation}

Similar expressions can be obtained for the direct CP asymmetry in \mbox{$B^+\to J/\psi K^+$} and the 
ratio \mbox{$\Xi\left(B^{\pm}\to J/\psi K^{\pm}, B_d\to J/\psi K^0\right)$} by making the substitution \begin{equation}
\tilde a_{\rm c} \to \epsilon \tilde a'_{\rm c}\:,\qquad \tilde \theta_{\rm c} \to \tilde \theta_{\rm c}'
+\pi\:,\qquad x \to \epsilon x'\:,\qquad \sigma \to \sigma' + \pi\:.
\end{equation}

\begin{figure}[t]
\center
\includegraphics[width=0.75\textwidth]{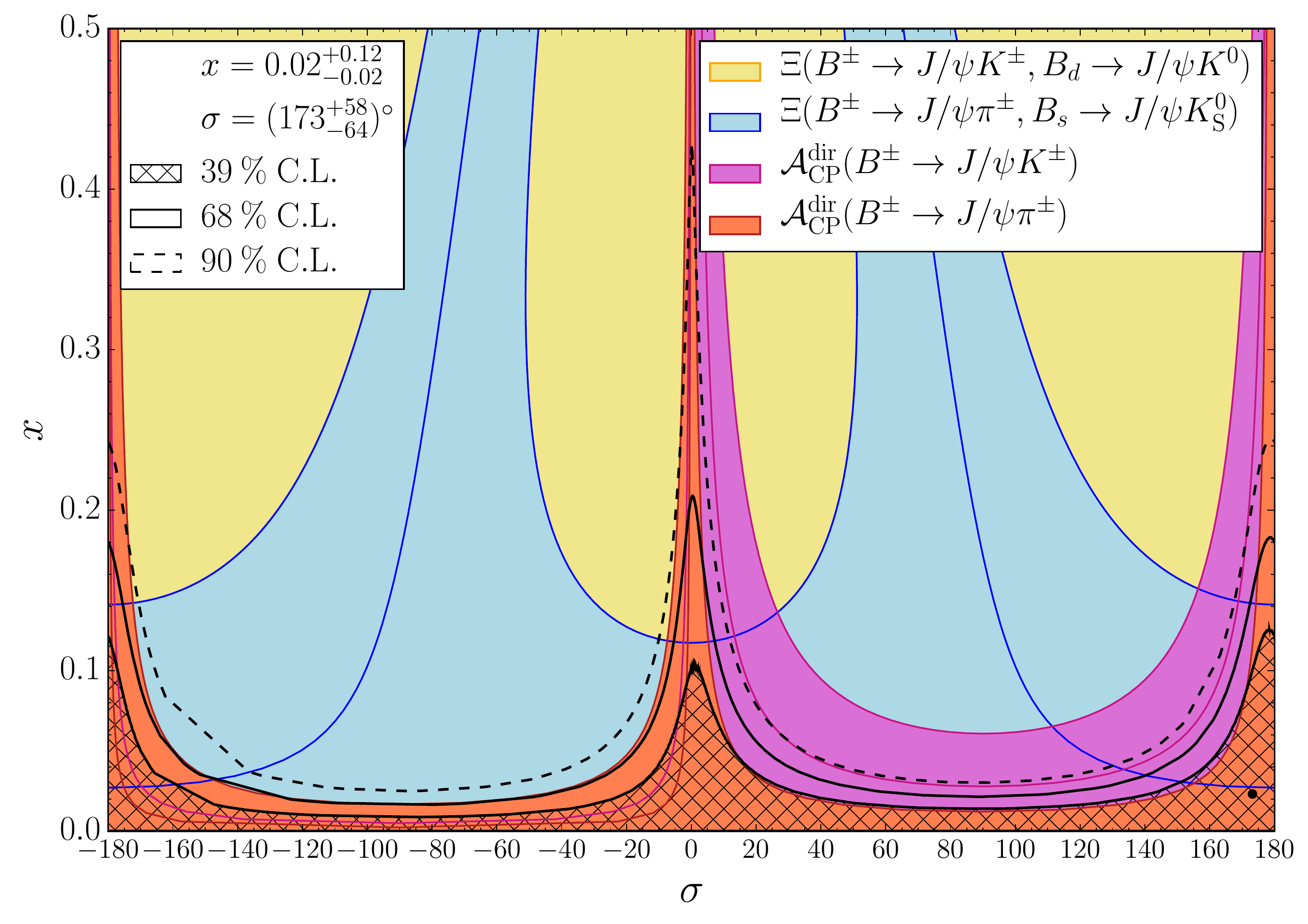}
\caption{Determination of the parameters $x$ and $\sigma$, which probe annihilation topologies
in $B^+\to J/\psi \pi^+$ and $B^+ \to J/\psi K^+$ decays, through intersecting 
contours corresponding to the current data for the CP asymmetries and branching ratio information. 
We show also the confidence level contours following from a $\chi^2$ fit.}
\label{Fig:Annihilation_Abs}
\end{figure}

Assuming
\begin{equation}
x'e^{i\sigma'} = x e^{i\sigma}
\end{equation}
and universal penguin parameters, i.e.\
\begin{equation}
\tilde a_{\rm c} e^{i\tilde \theta_{\rm c}} = \tilde a_{\rm c}' e^{i\tilde \theta_{\rm c}'}=
a e^{i\theta},
\end{equation}
the annihilation parameters $x$ and $\sigma$ can be obtained from a $\chi^2$ fit to the 
two direct CP asymmetries and the two $\Xi$ ratios listed above.
Including the observables $\gamma$ (from Eq.~\eqref{gamma-range}), $a$ and $\theta$ 
(from Eq.~\eqref{Eq:B2PV_chi2fit}) as Gaussian constraints results in the solution
\begin{equation}\label{Eq:Ann_chi2fit}
x = 0.02^{+0.12}_{-0.02} \:,\qquad \sigma = \left(173^{+58}_{-63}\right)^{\circ}\:,
\end{equation}
with the corresponding confidence level contours shown in Fig.~\ref{Fig:Annihilation_Abs}.
The result is compatible with $x=0$, which is consistent with our assumption to neglect contributions from 
annihilation topologies in the main $\chi^2$  fit.

The results in Eq.~\eqref{Eq:Ann_chi2fit} assume external input for the penguin parameters $a$ 
and $\theta$, and therefore do not take into account the back reaction of a non-zero value of
$xe^{i\sigma}$ on $ae^{i\theta}$. The annihilation topologies could lead to effects of similar size as
the exchange and penguin annihilation topologies, which can be probed through the
$B^0_s\to J/\psi \pi^0$ decay. In the future, with stringent constraints on the branching ratio of 
this channel, an extended fit could be made, including all additional topologies. But then we expect
to have also high-precision measurements of the CP violation in $B^0_s\to J/\psi K^0_{\rm S}$ 
available, allowing us to implement the strategy discussed in the main part of the paper. The extended
fit would nevertheless offer an interesting cross-check to complement the picture of the penguin 
parameters.

\newpage

\end{document}